\documentclass[11pt, oneside ]{article}

\usepackage[margin=1in]{geometry}  
\usepackage{color}            		
\usepackage{graphicx}				

\usepackage{amsmath,amssymb} 
\usepackage{bm}  
\usepackage{flafter}  
\usepackage{mathbbol}
\usepackage{txfonts}
\usepackage[T1]{fontenc}

\usepackage{aliascnt} 
\newtheorem{theorem}{Theorem}
\newaliascnt{definition}{theorem}
\newaliascnt{lemma}{theorem}
\newaliascnt{corollary}{theorem}
\newaliascnt{proposition}{theorem}
\newaliascnt{claim}{theorem}

\newtheorem{definition}[definition]{Definition}
\newtheorem{lemma}[lemma]{Lemma}
\newtheorem{corollary}[corollary]{Corollary}

\newtheorem{example}{Example}

\aliascntresetthe{definition}
\aliascntresetthe{lemma}
\aliascntresetthe{corollary}
\aliascntresetthe{proposition}
\aliascntresetthe{claim}

\newtheorem{remark}{Remark}

\newenvironment{proof}{%
  \noindent{\it Proof.\ }}{%
  \hspace*{\fill}\qed
  \vspace{2ex}\\}
\newenvironment{proofof}[1]{%
  \vspace{2ex}
  \noindent{\it Proof of #1.\ }}{%
  \hspace*{\fill}\qed
  \vspace{2ex}}
\usepackage[
pdftex,
bookmarks=true,
bookmarksnumbered=true,
bookmarkstype=toc,
colorlinks,
breaklinks,
pdfauthor=Seiichiro Tani,%
]{hyperref}
\definecolor{dullmagenta}{rgb}{0.4,0,0.4}   
\definecolor{darkblue}{rgb}{0,0,0.4}
\hypersetup{linkcolor=red,citecolor=blue,filecolor=dullmagenta,urlcolor=darkblue} 
\def\equationautorefname~#1\null{\textrm{Eq.~(#1)}\null}
\def\figureautorefname~#1\null{\textrm{Fig.~#1}\null}
\def\tableautorefname~#1\null{\textrm{Tab.~#1}\null}
\def\sectionautorefname~#1\null{\textrm{Sec.~#1\;}\null}
\def\subsectionautorefname~#1\null{\textrm{Sec.~#1\;}\null}
\def\subsubsectionautorefname~#1\null{\textrm{Sec.~#1\;}\null}
\def\pageautorefname~#1\null{\textrm{page~#1\;} \null}

\usepackage[ruled,lined]{algorithm2e}
\SetAlFnt{\small}
\SetAlCapFnt{\small}
\SetAlCapNameFnt{\small}
\SetAlCapHSkip{0pt}
\IncMargin{-\parindent}

\newcommand{\calS}{\mathcal{S}}
\newcommand{\calB}{\mathcal{B}}
\newcommand{\calF}{\mathcal{F}}

\newcommand{\abs}[1]{\lvert #1 \rvert}

\newcommand{\qed}{$\square$}

\newcommand{\closedset}[1]{[ #1]}
\newcommand{\cset}{\closedset}
\newcommand{\Closedset}[1]{\left[ #1 \right]}
\newcommand{\Cset}{\Closedset}
\newcommand{\deq}{:=}

\newcommand{\floor}[1]{\lfloor #1 \rfloor}

\newcommand{\ket}[1]{| #1 \rangle}
\newcommand{\Ket}[1]{\left| #1 \right\rangle}

\newcommand{\Openset}[1]{\left( #1 \right)}
\newcommand{\Oset}{\Openset}

\newcommand{\set}[1]{\{ #1 \}}
\newcommand{\Set}[1]{\left\{ #1 \right\}}
\newcommand{\true}{\mathsf{true}}
\newcommand{\false}{\mathsf{false}}
\renewcommand{\vec}[1]{\mathbf{#1}}

\newcommand{\Natural}{\mathbb{N}}

\newcommand{\entropy}[2]{\mathbf{H}_{#1}(#2)}
\newcommand{\Entropy}[2]{\mathbf{H}_{#1}\left(#2\right)}
\newcommand{\bentropy}[1]{\entropy{}{#1}}
\newcommand{\Bentropy}[1]{\Entropy{}{#1}}
\newcommand{\inner}[1]{\langle #1 \rangle}

\newcommand{\BDD}[2]{\calB(#1,#2)}
\newcommand{\BDDfp}{\calB(f,\pi)}
\newcommand{\cost}[3]{\mathsf{Cost}_{#1}(#2,#3)}
\newcommand{\costfp}[1]{\cost{#1}{f}{\pi}}
\newcommand{\var}{\mathsf{var}}
\newcommand{\child}[2]{\mathrm{child}_#2[#1]}
\renewcommand{\root}{\mathsf{r}}
\newcommand{\zero}{\mathsf{f}}
\newcommand{\one}{\mathsf{t}}

\newcommand{\Perm}[1]{\calS_{#1}}
\newcommand{\mincost}[1]{\textsc{mincost}_{#1}}

\newcommand{\tab}{\textsc{table}}
\newcommand{\node}{\textsc{node}}

\newcommand{\FS}[1]{\calF\calS(#1)}
\newcommand{\calg}{\mathsf{FS}}
\newcommand{\ccalg}{\mathsf{FS}^\ast}
\newcommand{\qalg}{\mathsf{OptOBDD}}
\newcommand{\cqalg}{\mathsf{OptOBDD}^\ast}
\newcommand{\qalgorithm}[1]{\qalg(#1)}
\newcommand{\cqalgorithm}[2]{\cqalg_{#1}(#2)}

\renewcommand{\vec}[1]{\mbox{\boldmath $#1$}}
\newcommand{\vb}{b}
\newcommand{\vx}{x}
\newcommand{\sizek}{\mathfrak{k}}
\newcommand{\sizei}{\mathfrak{i}}
\newcommand{\sizej}{\mathfrak{j}}

\author{Seiichiro Tani\\
NTT Communication Science Laboratories,\\
Nippon Telegraph and Telephone Corporation.\\
3-1, Morinosato-Wakamiya, Atsugi 243-0198, Japan.\\
\texttt{\small seiichiro.tani@acm.org}}
\date{}
\title{Quantum Algorithm for Finding the Optimal Variable Ordering for Binary Decision Diagrams%
\footnote{A preliminary version of this paper
appeared in the proceedings of 
the 17th Scandinavian Symposium and Workshops on Algorithm Theory ({SWAT})~\cite{Tan20SWAT}.
The journal version appears in Theoretical Computer Science, Vol. 1041, Article115230, 2025
(DOI: \href{https://doi.org/10.1016/j.tcs.2025.115230}{10.1016/j.tcs.2025.115230}).
}
\footnote{The author’s current aﬃliation is Waseda University, 1-6-1 Nishiwaseda, Shinjuku, Tokyo 169-8050, Japan.}}
\begin{document}
\sloppy
\begin{titlepage}
\thispagestyle{empty}
\maketitle
\begin{abstract}
An ordered binary decision diagram (OBDD) is 
a directed acyclic graph representing a Boolean function.
Since OBDDs have many nice properties as data structures,
they have been extensively studied for decades in theoretical and practical fields,
such as VLSI (Very Large Scale Integration) design, formal verification, machine learning, and combinatorial problems.
Arguably, the most crucial problem in using OBDDs is that
they may vary exponentially in size depending on their variable ordering (i.e., 
the order in which the variables are to be read) when they represent the same function.
Indeed, it is NP-hard to find an optimal variable ordering that minimizes an OBDD
for a given function.
Friedman and Supowit provided a clever deterministic algorithm with time/space complexity
$O^\ast(3^n)$, where $n$ is the number of variables of the function,
which is much better than the trivial brute-force bound $O^\ast(n!2^n)$.
This paper shows that a further speedup is possible with quantum computers
by presenting
 a quantum algorithm
that produces a minimum OBDD together with the corresponding variable ordering
in $O^\ast(2.77286^n)$ time and space 
with an exponentially small error probability. 
Moreover, this algorithm can be adapted to
constructing other minimum decision diagrams, such as zero-suppressed BDDs (ZBDDs or ZDDs).
\end{abstract}

\end{titlepage}


\pagestyle{plain} 
\section{Introduction}
\subsection{Background}
\begin{figure}[tb]
\centering
\includegraphics[width=10cm]{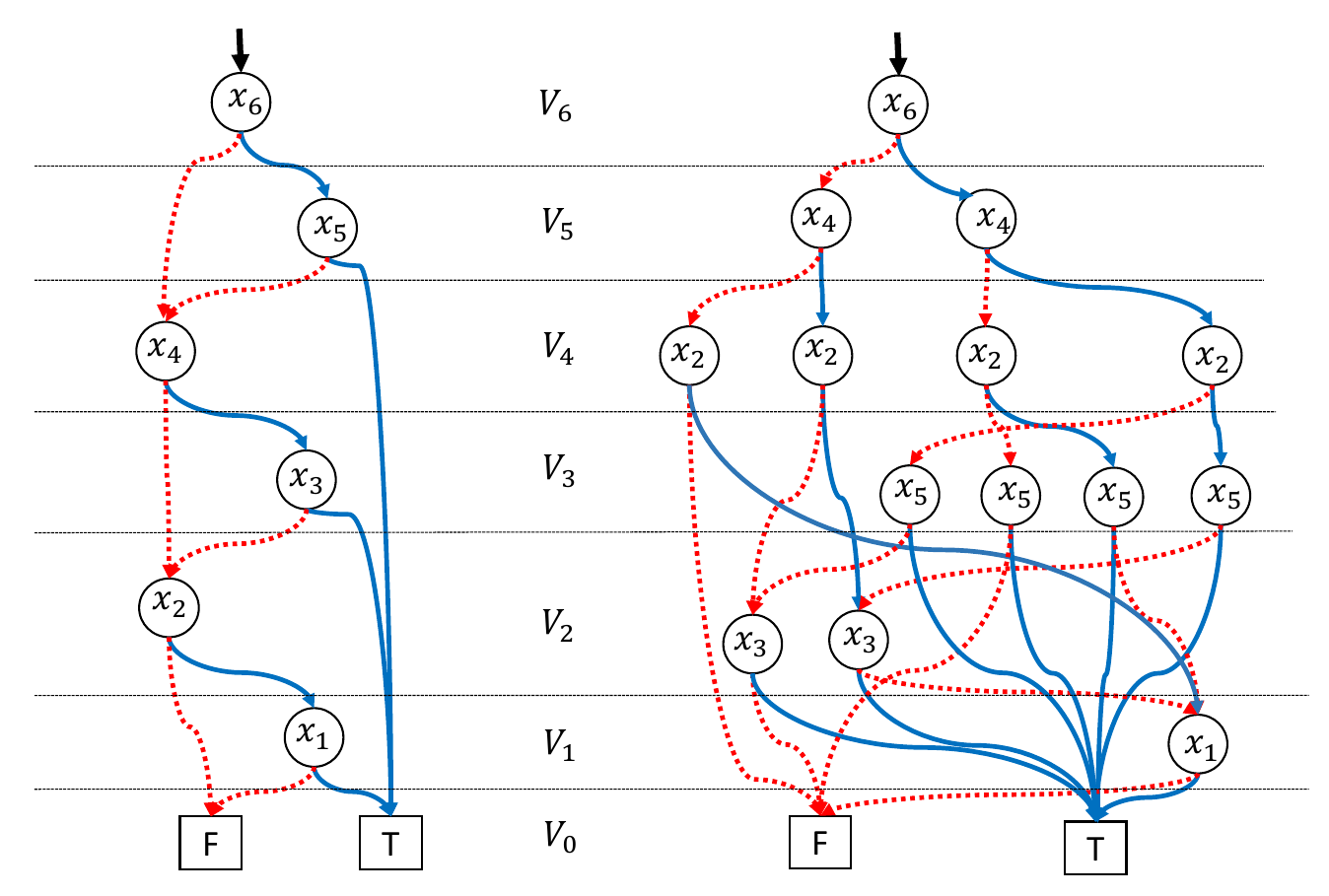}
\caption{The OBDDs represent the function $f(x_1,x_2,x_3,x_4,x_5,x_{6})=x_1\cdot x_2+x_3\cdot x_4+x_{5}\cdot x_{6}$  ("$\cdot$" and "$+$" denote AND and OR, respectively)
under two variable orderings: $(x_1,x_2,x_3,x_4,x_5, x_{6})$ (left) and $(x_1,x_3,x_5,x_2,x_4,x_6)$
(right), where the solid and dotted arcs express $1$-edges and $0$-edges, respectively,
and the terminal nodes for $\true$ and $\false$ are labeled with $\mathsf{T}$ and $\mathsf{F}$, respectively.
For each $n\in \Natural$, the function $f(x_1,\dots, x_{2n})=x_1\cdot x_2+x_3\cdot x_4+\dots+x_{2n-1}\cdot x_{2n}$
has  a $(2n+2)$-sized OBDD for the ordering $(x_1,\dots, x_{2n})$
and a $2^{n+1}$-sized OBDD for the ordering $(x_1,x_3,\dots, x_{2n-1},x_2,x_4,\dots,x_{2n})$~\cite[Sec.~8.1]{MeiThe98Book}. 
With the notations defined in \autoref{def:OBDD},
$\set{V_0,\dots, V_6}$ is the partition of the node set;
the top node is identified with $\root$; and
the bottom two nodes are identified with $\one$ and $\zero$.
A more detailed explanation of these OBDDs is provided in 
\autoref{example:OBDDs}.}
\label{fig:OBDDs}
\end{figure}

\paragraph*{Ordered binary decision diagrams.} 
The ordered binary decision diagram (OBDD)
is one of the data structures
that have been most often used for decades to represent Boolean functions in practical situations,
such as VLSI design, formal verification, optimization of combinatorial problems, and machine learning,
and it has been extensively studied from both theoretical and practical standpoints (see standard textbooks and surveys, e.g., Refs.~\cite{Bry92CSUR,MeiThe98Book,DreBec98Book,Weg00Book,Knu09Book,Bry18BookChapter}).
Moreover, many variants of OBDDs have been invented to more efficiently represent data with properties observed frequently in specific applications (e.g., Refs.~\cite{Min93DAC,BryChe95DAC,BahFroGaoHacMacParSom97FMSD,ClaMcmZhaFujYan97FMSD,Min11SAT}).
More technically speaking, OBDDs are 
directed acyclic graphs representing Boolean functions
and also known as  special cases of oblivious read-once branching programs
studied in the field of complexity theory. 
The reason for OBDDs' popularity lies in their nice properties ---
they can be uniquely determined up to isomorphism for each function
once \emph{variable ordering} (i.e., the order in which to read the variables) is fixed
and, thanks to this property, 
the equivalence of functions can be checked by just testing the isomorphism between
the OBDDs representing the functions (this can be done in linear time since each OBDD is a directed acyclic graph with a single source node
and labeled edges).
In addition, binary operations such as \textsc{AND} and \textsc{OR} between two functions can be performed efficiently 
over the OBDDs representing those functions~\cite{Bry86IEEETC}.
Since these properties are essential in many applications, OBDDs have gathered much
attention from various research fields.
To enjoy these nice properties, however, we need to address a crucial problem,
which is that
OBDDs may vary exponentially in size depending on their {variable ordering}.
For instance, a Boolean function 
$f(x_1,\dots, x_{2n})=x_1\cdot x_2+x_3\cdot x_4+\dots+x_{2n-1}\cdot x_{2n}$ ("$\cdot$" and "$+$" denote AND and OR, respectively)
has a $(2n+2)$-sized OBDD for the ordering $(x_1,\dots, x_{2n})$
and a $2^{n+1}$-sized OBDD for the ordering $(x_1,x_3,\dots, x_{2n-1},x_2,x_4,\dots,x_{2n})$~\cite[Sec.~8.1]{MeiThe98Book} (see \autoref{fig:OBDDs}
for the case where $n=3$).
This example is not rare; it could happen in many concrete functions one encounters.
Thus, since the early stages of OBDD research, one of the most central problems has been finding an optimal variable ordering, i.e.,
one that minimizes OBDDs.
Since there are $n!$ permutations over $n$ variables, the brute-force search
requires at least $n!=2^{\Omega(n\log n)}$ time to find an optimal variable ordering.
Indeed, finding an optimal variable ordering for a given function is an NP-hard problem
(see~\autoref{sec:RelatedWork} for the studies on the hardness).

To tackle this high complexity,
many heuristics have been proposed to find an optimal variable ordering or 
a relatively good one.
These heuristics 
work well for Boolean functions appearing
in specific applications
since they are based on very insightful observations,
but they do not guarantee a worst-case time complexity
lower than that achievable with the brute-force search.
The only algorithm with a much lower worst-case time complexity bound, 
$O^\ast(3^n)$  time ($O^\ast(\cdot)$ hides a polynomial factor),
than the brute-force bound $O^\ast(n!2^n)$ for \emph{all} Boolean functions with $n$ variables was provided
by Friedman and Supowit~\cite{FriSup90IEEETC}, and that was over thirty years ago!

In practice, constructing a \emph{minimum} OBDD is often too costly.
Moreover, even if the optimal variable ordering is found,
it may become non-optimal when the function changes during processing a task by imposing additional constraints.
Nevertheless, theoretically sound methods for finding an optimal variable ordering
are worth studying for several reasons, such as to judge the optimization
quality of heuristics and to be able to apply such methods at least to parts of the OBDDs
within a heuristics procedure~\cite[Sec.~9.22]{MeiThe98Book}.

\paragraph{Quantum Speedups of Dynamic Programming}
Grover's quantum search algorithm~\cite{Gro96STOC} and its variants achieve quadratic speedups over
any classical algorithm for \emph{unstructured search}, a very fundamental problem (effectively, exhaustive search is the only classical strategy).
Thus, one of the merits of the quantum search is its broad applicability.
However, it does not immediately mean quantum speedups for
all problems to which the quantum search is applicable,
since 
better classical algorithms may exist than the simple exhaustive search.
Indeed, quantum search for an optimal variable ordering of the OBDD from among $n!$ candidates
takes approximately $\sqrt{n!}\approx 2^{\frac{1}{2}n\log n}$ time. In contrast, the best classical algorithm
takes only $O^\ast(3^n)=O^\ast(2^{(\log_2 3)n})$.
These classical algorithms often employ powerful algorithmic techniques 
such as dynamic programming, divide-and-conquer, and branch-and-bound.
One typical strategy to gain quantum speedups would be
to find parts of exhaustive search (often implicitly) performed within such classical algorithms
and apply the quantum search to those parts.
For instance, D\"urr et al.~\cite{DurHeiHoyMha06SICOMP} provided quantum algorithms for some graph problems,
among which the quantum algorithm for the single-source shortest-path problem 
achieves a quantum speedup
by applying a variant of Grover's search algorithm to select the cheapest border edge in Dijkstra's algorithm.
However, applying the quantum search in this way
does not work
when the number of states in a dynamic programming algorithm
is much larger than the number of predecessors of each state.
For instance, the Traveling Salesman Problem (TSP) can be solved
in $O^\ast(2^n)$ time by a classical dynamic programming algorithm,
but locally applying the quantum search can attain at most a polynomial-factor improvement.
Recently, Ambainis et al.~\cite{AmbBalIraKokPruVih19SODA} has introduced break-through techniques 
to speed up dynamic programming approaches.
They provide quantum algorithms that solve a variety of vertex ordering problems on graphs
in $O^\ast(1.817^n)$ time, graph bandwidth in  $O^\ast(2.946^n)$ time, and
TSP and minimum set cover in $O^\ast(1.728^n)$ time, where $n$ is the number of vertices in the graphs.

\subsection{Our Results}
In this paper, 
we show that quantum speedup is possible for the problem of finding an optimal variable ordering of the OBDD for a given function. 
This is the first quantum speedup for OBDD-related problems.
Our algorithms assume the quantum random access memory (QRAM) model~\cite{GioLloMac08PRL}, commonly used in the literature concerning quantum algorithms.
In the model, one can read contents from or write them into quantum memory in a superposition with a logarithmic factor overhead in the time complexity.
We provide our main result in the following theorem.
\begin{theorem}\label{th:main_informal}
There exists a quantum algorithm that,
for a function $f\colon \set{0,1}^n\to\set{0,1}$ given as its truth table,
produces a minimum OBDD representing $f$
together with the corresponding variable ordering
in $O^\ast(\gamma^n)$ time and space
with an exponentially small error probability with respect to $n$, 
where the constant $\gamma$ is
at most $2.77286$.
Moreover, the OBDD produced by the algorithm is always a valid one
for $f$,
although it is not minimum with an exponentially small probability.
\end{theorem}

This improves upon the classical best bound 
$O^\ast(3^n)$~\cite{FriSup90IEEETC} on time/space complexity.
The classical algorithm achieving this bound
is a deterministic one. 
However, there are no randomized algorithms
that compute an optimal variable ordering in asymptotically less time complexity
as far as we know.

It may seem somewhat restricted to assume that the function $f$ is given as its truth table,
since there are other common representations of Boolean functions
such as DNFs, CNFs, Boolean circuits, and OBDDs.
However, this is not the case.
Our algorithm works without increasing the order of complexity%
\footnote{
Here, we consider the complexity with respect to \emph{the number of variables}. The complexity with respect to \emph{input size} may vary drastically.} in more general settings
where the input function $f$ is given as any representation
such that the value of $f$ on any specified assignment can be computed
over the representation in polynomial time in $n$, such as
polynomial-size DNFs/CNFs/circuits and 
OBDDs of any size.%
\footnote{Regardless of the size of input OBDD, the value of $f$ on any specified assignment can be computed in $O(n)$ time by traversing the path corresponding to the assignment from the root.}
This is because, in such cases, the truth table of $f$ can be prepared
in $O^\ast(2^n)$ time/space, which is negligible compared with the total time/space complexity, and the minimum OBDD is computable from that truth table
with our algorithm. We restate 
\autoref{th:main_informal}
in a more general form as follows.

\begin{corollary}\label{th:quantum_algorithm_composition:oracle}
Let $R(f)$ be any representation of a Boolean function $f$ with $n$ variables
such that the value of $f$ on any given assignment $x\in \set{0,1}^n$ can be computed
on $R(f)$ in polynomial time with respect to $n$.
Then, there exists a quantum algorithm that,
for a function $f\colon \set{0,1}^n\to\set{0,1}$ given as $R(f)$,
produces a minimum OBDD representing $f$
together with the corresponding variable ordering
in $O^\ast(\gamma^n)$ time and space
with an exponentially small error probability with respect to $n$, 
where the constant $\gamma$ is
at most $2.77286$. 
Possible representations as $R(f)$ include
polynomial-size DNFs/CNFs/circuits and OBDDs of any size for the function $f$.
\end{corollary}

There are many variants of OBDDs, among which the zero-suppressed BDDs (ZDDs or ZBDDs) introduced by Minato~\cite{Min93DAC} are very powerful in dealing with combinatorial problems (see Knuth's famous book~\cite{Knu09Book} for how to apply ZDDs to such problems). 
With slight modifications, 
our algorithm can construct a minimum ZDD with the same time/space complexity. 
We believe that similar speedups are possible for many other variants of OBDDs
(adapting our algorithm to multiterminal BDDs (MTBDDs)~\cite{BahFroGaoHacMacParSom97FMSD,ClaMcmZhaFujYan97FMSD}
is almost trivial).

\subsection{Technical Outline}
The first step to take is to somehow adapt the dynamic programming approach of the classical algorithm~\cite{FriSup90IEEETC} (called $\calg$) to the framework provided by Ambainis et al.~\cite{AmbBalIraKokPruVih19SODA}.
Consider a Boolean function $f$ over $n$ variables: $x_1,\dots, x_n$.
Roughly speaking, $\calg$ determines the variable ordering of the minimum OBDD
for $f$ by performing dynamic programming from
the variable to be read last toward that to be read first.
More concretely, let $(x_{\pi[1]},\dots, x_{\pi[n]})$ be the variable ordering 
from the one read last ($x_{\pi[1]}$) to the one read first ($x_{\pi[n]}$),
where $\pi=(\pi[1],\dots,\pi[n])$ is a permutation over $[n]\deq\set{1,\dots, n}$.
For $\sizek=1,\dots, n$ in this order, and for every subset $K\subseteq [n]$
of cardinality $\sizek$, 
the algorithm $\calg$ computes the minimum size (denoted by $\mincost{K}$) of the bottom $\sizek$ layers%
\footnote{
Strictly speaking, the minimum size is the sum of the nodes
in the layers associated with bottom $\sizek$ variables \emph{and} Boolean values $\true,\false$
(e.g., $V_0,\dots, V_{\sizek}$ in \autoref{fig:OBDDs}). For simplicity, however, we do not explicitly count the number of nodes associated with Boolean values in this subsection.} 
of the OBDD under the restriction of $\set{\pi[1],\dots,\pi[\sizek]}=K$
from the minimum size (denoted by $\mincost{K\setminus \set{k}}$) of the bottom $\sizek -1$ layers  of the OBDD
under the restriction of 
 $\set{\pi[1],\dots,\pi[\sizek -1]}=K\setminus \set{k}$
for each $k\in K$.
Thus, by thinking of each node $z\in \set{0,1}^n$ of Hamming weight $\sizek$ in a Boolean hypercube as the characteristic vector of $K$, 
the algorithm $\calg$ can be seen as solving a kind of shortest path problem on a Boolean hypercube with dynamic programming from node $0^n$ to node $1^n$.
Hence, Ambainis et al.'s framework seems applicable.
Their results depend on the property that
a large problem can be divided into the same kind of subproblems or,
in other words, 
scale-down versions of the original problem
in the sense that they can be solved with the same algorithm.
This property naturally holds in many graph problems.
In our case, firstly, it is unclear whether the problem can be divided into subproblems.
Secondly, even if it is possible,
the subproblems would be
to optimize the ordering of variable 
starting from the middle variable or even from the opposite end, i.e., from the variable to be read \emph{first},
toward the one to be read \emph{last}.
Such subproblems cannot be solved with the algorithm $\calg$, and,
in particular,
optimizing in the latter case
essentially
requires the equivalence check of subfunctions of $f$,
which is very costly. 

Our technical contribution is 
to find, 
by carefully observing the unique properties of OBDDs,
that it is possible to (even recursively) divide the original problem into
scale-down versions of a \emph{generalization} of the original problem,
to generalize the algorithm $\calg$ so that it can solve the subproblems,
and to use the quantum minimum finding algorithm in order to efficiently select 
the subproblems that essentially contribute to the optimal variable ordering.

More concretely, fix any $\sizek \in [n]$.
For a subset $K\subseteq [n]$ with cardinality $\sizek$,
let $\mincost{\inner{K,[n]\setminus K}}$ be the minimum size of the OBDD 
over all variable ordering $\pi$ under the restriction of 
$\set{\pi[1],\dots,\pi[\sizek]}=K$.
It turns out in \autoref{lm:divide-and-conquer} that
the minimum size of the OBDD
is the minimum of $\mincost{\inner{K,[n]\setminus K}}$
over all $K\subseteq [n]$ of cardinality $\sizek$.
Then, by computing 
$\mincost{\inner{K,[n]\setminus K}}$ for every $K$ of cardinality $\sizek$
and taking the minimum, we can obtain the minimum size of the OBDD
(and the associated variable ordering by computing the variable ordering $\pi$ with
$\set{\pi[1],\dots,\pi[\sizek]}=K$
while computing $\mincost{\inner{K,[n]\setminus K}}$).
Intuitively, we compute the shortest path going through each node 
on the plane composed of the nodes with weight $\sizek$
in the Boolean hypercube, and choose the best one.
Since the shortest path from node $0^n$ to node $1^n$ must go through one of such nodes,
this gives the desired solution.
If we perform this computation classically, 
it takes $\binom{n}{\sizek}$ times the time required to compute 
$\mincost{\inner{K,[n]\setminus K}}$ for a single $K$, i.e.,
\[
\binom{n}{\sizek}\cdot T\left(\mincost{\inner{K,[n]\setminus K}}\right),
\]
where $T(h)$ is is  the time required to compute $h$ (we also use this notation 
below in this subsection).

The fundamental idea for quantum speedup
is to compute
$\mincost{\inner{K,[n]\setminus K}}$ 
for every $K$ of cardinality $\sizek$ \emph{in superposition}
and find a specific $K$ that achieves the minimum of $\mincost{\inner{K,[n]\setminus K}}$ 
(together with the associated variable ordering) 
by using the quantum minimum finding algorithm (\autoref{lm:QuantumMinimumFinding}).
This takes the time of
\[
O\left(\sqrt{\binom{n}{\sizek}}\right)\cdot T\left(\mincost{\inner{K,[n]\setminus K}}\right).
\]
This algorithm outputs the correct answer for every choice of $\sizek$.
However, the complexity depends on the choice of $\sizek$. Thus, we must choose $\sizek$ carefully to minimize the complexity.

To compute $\mincost{\inner{K,[n]\setminus K}}$, we divide it
into the following two parts:
\begin{itemize}
\item
the minimum size $\mincost{K}$
of the bottom $\sizek$-layers of the OBDD, 
under the restriction of $\set{\pi[1],\dots,\pi[\sizek]}=K$,
\item
the the minimum size of the upper $(n-\sizek)$-layers of the OBDD
under the restriction of $\set{\pi[\sizek+1],\dots,\pi[n]}=[n]\setminus K$.
\end{itemize}
Intuitively, the shortest path going through the node $K$
in the Boolean hypercube can be divided into the shortest path
from node $0^n$ to node $K$
and the shortest one from node $K$ to node $1^n$.
It turns out that $\calg$ can be generalized to compute these two parts \emph{sequentially}
(i.e., to compute the first part and then the second part using the information of the first part),
which implies 
\[
T\left(\mincost{\inner{K,[n]\setminus K}}\right)
=T\left(\mincost{K}[K]\right)+T\left(\mincost{\inner{K,[n]\setminus K}}[[n]\setminus K]\right),
\]
where $\mincost{K}[K]$ and
$\mincost{\inner{K,[n]\setminus K}}[[n]\setminus K]$ represent
the former and the latter parts, respectively.

To obtain a better bound, 
we use a similar idea (with some modifications) to compute
$\mincost{K}[K]$. 
More concretely, 
for a subset $K'\subseteq K$ with cardinality 
$\sizek' \in [\sizek]$, 
let $\mincost{\inner{K',K\setminus K'}}$ be the minimum size of the 
bottom $\sizek'$ layers of the OBDD
over all variable ordering $\pi$ under the restriction of 
$\set{\pi[1],\dots,\pi[\sizek']}=K'$ and $\set{\pi[1],\dots,\pi[\sizek]}=K$.
Then, we have
\[
T\left(\mincost{K}[K]\right)=O\left(\sqrt{\binom{\sizek}{\sizek'}}\right)\cdot T\left(\mincost{\inner{K',K\setminus K'}}\right).
\]

Repeating this composition recursively
is essentially equivalent to  
considering $d$  division points $ \sizek_1, \dots , \sizek_d$ ($0< \sizek_1< \dots< \sizek_d<n$)
and optimizing each of the $(d+1)$ suborderings $(\pi[1],\dots, \pi[\sizek_1]), 
(\pi[\sizek_1+1],\dots, \pi[\sizek_2]), \dots, (\pi[\sizek_d+1],\dots, \pi[n])$.
Intuitively, 
we consider $d$ planes in the Boolean hypercube,
where they consist of the points with weights
$ \sizek_1, \dots , \sizek_d$, respectively,
compute the shortest path among the paths going through
a fixed point on each of the $d$ planes.
Then, we choose the shortest one over all possible combinations
of the fixed points on the planes.
The last step is implemented as (the recursion of) the quantum minimum finding
over all nodes on each plane.

We obtain the final algorithm by 
optimizing  $\sizek_1, \dots, \sizek_d$, depending on $d$.
Our complexity can be obtained for a specific $d$ at which
the complexity is almost saturated (while larger $d$ generally yields better complexity).

\subsection{Related Work}\label{sec:RelatedWork}
The studies related to minimizing OBDDs are so numerous
that we cannot cover all of them. 
We thus pick up some of the purely theoretical work.

Meinel and Slobodov{\'a}~\cite{MeiSlo94MFCS} proved that it is NP-hard to construct an optimal OBDD for a Boolean function given by a logical circuit, a DNF, a CNF, or an OBDD, even if the optimal OBDD is of constant size.
Tani, Hamaguchi and Yajima~\cite{TanHamYaj96IEICE} proved
that it is NP-hard to 
improve the variable ordering (and thus, to find an optimal variable ordering)
for a given \emph{multi}-rooted OBDD, where the NP-hardness is proved by a reduction from an
NP-complete problem, Optimal Linear Arrangement~\cite{GarJoh79Book}.
Bollig and Wegener~\cite{BolWeg96IEEETC} finally proved
the NP-hardness for a given \emph{single}-rooted OBDD by providing a sophisticated reduction from the same problem.
This is still true if the input function is restricted to monotone functions~\cite{IwaNouYaj98}.
Minimizing the width of an OBDD is also NP hard~\cite{Bol16ToCS}.
As for approximation hardness, Sieling~\cite{Sie02DAM,Sie02InfoComp} proved that
if there exists a polynomial-time approximation scheme for
computing the size of the minimum OBDD for a given OBDD,
it then holds that $\mathrm{P}=\mathrm{NP}$.

It would be nice if, for every function,
there exists at least one variable ordering
under which the OBDD for the function is of a size bounded by a polynomial.
As one may expect, 
this is not the case.
It can be proved by a counting argument that
there exists a (family of) functions for which the OBDD size
grows exponentially in the number of variables
under \emph{any} variable ordering~\cite{Lee59BSTJ,HehChe92IEEETC,HeaMer94IEEETC}.
Moreover, concrete examples of such functions are known:
the multiplication function~\cite{Bry91IEEETC},
a threshold function~\cite{HosTakKanYaj97TCS},
and the division function~\cite{HorYaj97ISAAC}
(for other classes of Boolean functions, see Ref.~\cite{SawTakYaj94IEICE,Hea93JET,HeiMol00IEEETC}).
The OBDD size is also studied from the viewpoint of 
computational learning and knowledge-bases~\cite{TakYaj00DAM,HorIba02AI}.

In applying OBDDs to graph problems, it is possible to find
variable orderings for which OBDD size is nontrivially upper-bounded
in terms of specific measures characterizing graph structures~\cite{TanIma94ISAAC,SekImaTan95ISAAC}.
A similar concept was discovered for ZDDs~\cite{Min93DAC} by Knuth~\cite{Knu09Book}.
This concept is now called the \emph{frontier method},
and many works are based on it.

\subsection{Organization}
\autoref{sec:preliminaries} defines basic notations
and ordered binary decision diagrams (OBDDs), then reviews the algorithm
invented by Friedman and Supowit, and finally provides
relevant basics of quantum computing.
\autoref{sec:divide-and-conquer} presents a basic version of our algorithm based on divide-and-conquer techniques.
\autoref{sec:composition} provides 
a recursive application of the basic version
as the final version of our algorithm.
Finally, \autoref{sec:conclusion} summarizes our results and discusses open questions.

\section{Preliminaries}\label{sec:preliminaries}
\subsection{Basic Terminology}
Let $\Natural$ be the set of natural numbers.
For each $n\in \Natural$, let $[n]$ be the set $\set{1,\dots, n}$,
and $\Perm{n}$ be the permutation group over $[n]$.
A singleton set $\set{i}$ may be denoted by $i$ for notational simplicity if it is clear from the context;
for instance, $I\setminus \set{i}$ may be denoted by $I\setminus i$, if we know that $I$ is a set.
For any subset $I\subseteq [n]$, let
$\Pi_n(I)$ be the set of $\pi\in \Perm{n}$ such that the first $\sizei\deq \abs{I}$ members 
$\set{\pi[1],\dots,\pi[\sizei]}$ constitutes $I$, i.e., 
\[
\Pi_n(I)\deq \Set{\pi\in \Perm{n} \colon \set{\pi[1],\dots,\pi[\sizei]}=I}\subseteq\Perm{n}.
\]
For simplicity, we omit the subscript $n$ and write $\Pi(I)$.
More generally, for any two disjoint subsets $I,J\subseteq[n]$ with $\sizei\deq \abs{I}$ and $\sizej\deq \abs{J}$, let
\[
\Pi_n(\inner{I,J})\deq \Set{\pi\in \Perm{n} \colon \set{\pi[1],\dots,\pi[\sizei]}=I, \set{\pi[\sizei+1],\dots,\pi[\sizei+\sizej]}=J}\subseteq\Perm{n}.
\]
For any disjoint subsets $I_1, \dots,I_m\subseteq [n]$ for $m\in [n]$, $\Pi_n(\inner{I_1,\dots, I_m})$ is defined similarly.
For simplicity, we may allow $I$ to denote $\inner{I}$, if it is clear from the context.

The union operation over \emph{disjoint} sets may be denoted by $\sqcup$ (instead of $\cup$)
to emphasize the disjointness of the sets.

For $n$ Boolean variables $x_1,\dots, x_n$, any set $I\subseteq [n]$ with $\sizei\deq \abs{I}$,
and any Boolean vector $\vb=(b_1,\dots, b_{\sizei})\in \set{0,1}^{\sizei}$,
$\vx_I$ denotes the ordered set $(x_{j_1},\dots, x_{j_{\sizei}})$,
where 
$\set{j_1,\dots,j_{\sizei}}=I$ 
and $j_1<\dots <j_{\sizei}$,
and $\vx_I=\vb$ denotes $x_{j_i}=b_i$ for each $i\in [\sizei]$.
For any Boolean function $f\colon \set{0,1}^n \to \set{0,1}$ with variables $x_1,\dots, x_n$,
$f|_{\vx_I=\vb}$ denotes the function 
obtained by restricting $f$ with $\vx_I=\vb$.
If $I$ is a singleton set, say, $I=\set{i}$, 
we may write $\vx_i$ and $f|_{\vx_i=b}$ to mean $x_{\set{i}}$ and $f|_{\vx_{\set{i}}=b}$, respectively,  for notational simplicity. We say that $g$ is a \emph{subfunction} of $f$ if $g$ is equivalent to the function $f|_{\vx_I=\vb}$
for some $I\subseteq [n]$ and $\vb\in \set{0,1}^{\sizei}$.

For any function $g(n)$ in $n$, we use the notation $O^\ast(g(n))$ to hide a polynomial factor in $n$.
Furthermore,  $X\lessapprox Y$ denotes $X=O^\ast(Y)$.

We use the following upper bound many times in this paper.
For $n\in \Natural$ and $\ell\in [n] \cup \set{0}$,
it holds that
$
\binom{n}{\ell}\le 2^{n\, \bentropy{\ell/n}},
$
where $\bentropy{\cdot}$ represents the binary entropy function $\bentropy{\delta}\deq -\delta \log_2 \delta -(1-\delta) \log_2 (1-\delta)$ for any $\delta\in [0,1]$. 
Note that
$\binom{\floor{\beta n}}{\floor{\alpha n}}=O(2^{\beta n \bentropy{\alpha/\beta}})$
for constants $0< \alpha<\beta\le 1$
since it holds
by Taylor expansion
that
$\abs{\bentropy{b}-\bentropy{a}}=O(1/n)$ 
if two values $a,b$ satisfy that $\epsilon<a<b<1$ for any constant $\epsilon>0$ and 
$b-a=O(1/n)$.

To help intuitively understand the role of each variable, we consistently use different letters as follows:
a set is denoted by an uppercase roman letter (e.g., $I,J,K$),
an element of a set
by a lowercase roman letter (e.g, $i\in I$, $j\in J$, $k\in K$),
the cardinality of a set by a Fraktur (e.g., $\sizei$ for $|I|$, $\sizej $ for $\abs{J}$, and 
$\sizek$ for $\abs{K}$).
\subsection{Ordered Binary Decision Diagrams}\label{sec:OBDD}
We provide a quick review of OBDDs.
For more details, 
consult standard textbooks or survey papers
(e.g., Refs.~\cite{Bry92CSUR,MeiThe98Book,DreBec98Book,Weg00Book,Bry18BookChapter}).

An OBDD is a special case of read-once oblivious branching programs in complexity-theoretic terms
that is, branching programs satisfying the following conditions:
each variable is read at most once
on each directed path from the root to a terminal node, and
the orderings of variables to be read on all such paths are consistent with a certain fixed ordering.
In other words, an OBDD for a Boolean function $f$ is a directed acyclic graph that represents Shannon expansion of $f$ according to a fixed ordering of variables, where each node of the OBDD corresponds to a subfunction obtained by Shannon expansion of $f$. Hence, the number of nodes in the OBDD is at most $\frac{2^n}{n}(1+o(1))$~\cite[Exercise 6.1]{AroBar09Book}, where $n$ is the number of variables of $f$.

Before providing a formal definition of OBDDs, it would help understand to see examples of OBDDs in \autoref{fig:OBDDs}.
Each path from the topmost node to one of the bottom nodes corresponds to a set of assignments to the variables in such a way that
the 0/1 value attached to each directed edge on the path
indicates the Boolean value that is assigned to the variable attached to the source node of the edge.
The bottom node at which the path terminates is labeled with the function value 
determined by any one of the assignments in the set (since the definition of OBDDs guarantees that the function values on all those assignments are the same).
In the left OBDD, for instance, the path consisting of only dotted (red) edges corresponds to
the set of the eight assignments satisfying that $x_2=x_4=x_6=0$, in which case the function value is $\false$
as the path terminates at the bottom node labeled with $\mathsf{F}$ (indicating $\false$).

\begin{figure}[t] 
   \centering
   \includegraphics[width=15cm]{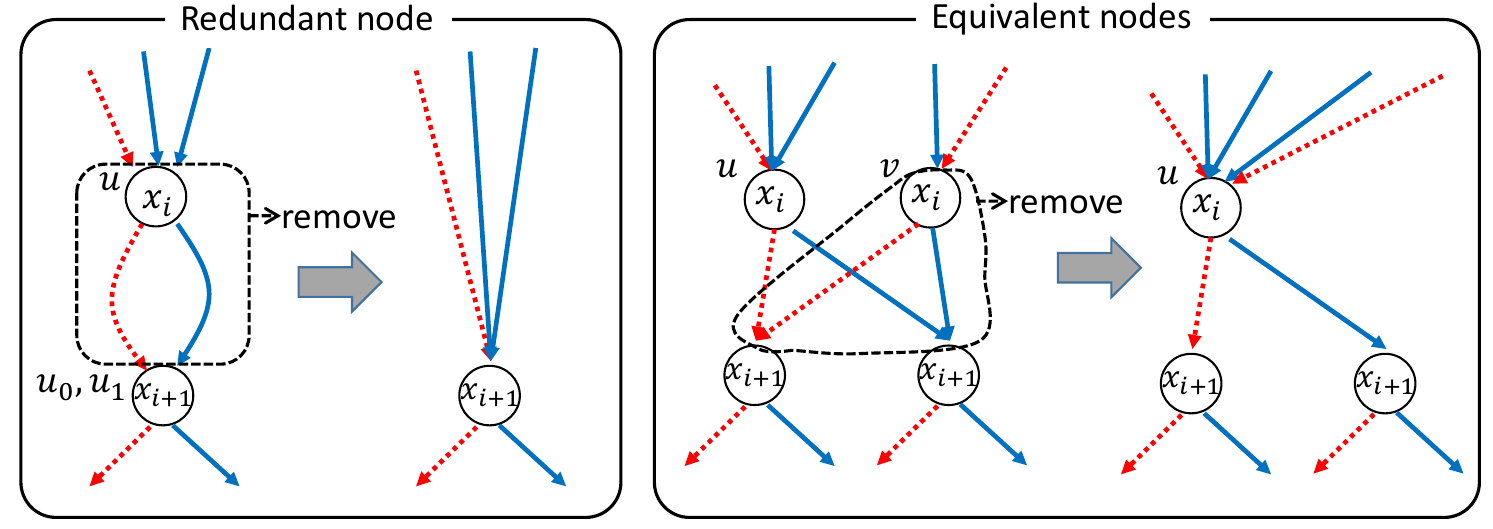} 
   \caption{Examples of a redundant node (left) and a pair of equivalent nodes (right), and their respective removal rules,  where the solid and dotted arcs express $1$-edges and $0$-edges, respectively. In the example of a redundant node, $u_0$ and $u_1$ denote $\child{u}{0}$ and $\child{u}{1}$, respectively.}
   \label{fig:rules}
\end{figure}

\begin{definition}[OBDDs]\label{def:OBDD}
For any Boolean function $f\colon \set{0,1}^n \to \set{0,1}$ over variables $x_1,\dots, x_n$ and any permutation $\pi\in \Perm{n}$ (called a \emph{variable ordering}),
an OBDD $\BDD{f}{\pi}$ is a single-rooted directed acyclic graph $G(V,E)$ defined as follows (see also \autoref{example:OBDDs} below).

\begin{enumerate}
\setlength{\itemsep}{0cm}
\item 
The node set $V$ is the union of two disjoint sets $N$ and $T$ of 
\emph{non-terminal} nodes with out-degree two and \emph{terminal} nodes with out-degree zero, respectively, where $T$ contains exactly two nodes:
$T=\set{\zero,\one}$.
The set $N$ contains a unique source node $\root$ (i.e., the node with in-degree zero) called the \emph{root}.

\item 
$\BDD{f}{\pi}$ is a leveled graph with $n+1$ levels.
Namely, the node set can be partitioned into $n$ subsets:
$
V\deq V_0\sqcup V_1\sqcup \dots \sqcup V_n,
$
where $V_n\deq \set{\mathsf{r}}$ and 
$V_0\deq T=\set{\mathsf{t},\mathsf{f}}$,
such that
each directed edge $(u,v)\in E$
is in $V_\ell\times V_m$
for a pair $(\ell,m)\in [n]\times (\set{0}\sqcup [n-1])$
with $\ell>m$.
For each $\ell\in[n]$, subset $V_\ell$
(called the \emph{level ${\ell}$})
is
associated with the variable $x_{\pi[\ell]}$
in the sense that each node in $V_\ell$ is labeled with $x_{\pi[\ell]}$.%
\footnote{In the standard definition, 
the variable ordering is defined as
the ordering in which the variables to be read, that is, 
the \emph{reverse} ordering of $\pi$. 
Our definition
follows the one given in~\cite{FriSup90IEEETC}
to avoid complicated subscripts of variables
in describing the algorithms in the following section.}
For convenience, we define the function $\var\colon N \to [n]$
that maps each non-terminal node 
to the index  of the associated variable,
so that
$\var(u)=\pi[\ell]$
for $u\in V_\ell \ (\ell\in [n])$.

\item The two edges emanating from every non-terminal node are called the \emph{$0$-edge} and the \emph{$1$-edge}, labeled with 0 and 1, respectively. 
 For every non-terminal node $u\in N$, let $\child{u}{0}$ and $\child{u}{1}$ be the destinations of the $0$-edge and $1$-edge of $u$, respectively.

\item 
Let $\calF(f)$ be the set of all subfunctions of $f$.
Define a bijective map $F\colon V\to \calF(f)$ as follows:
(a) 
$F(\root)=f$ for $\root\in V_n$;
(b) $F(\one)=\true$ and $F(\zero)=\false $ for $\one,\zero\in V_0$;
(c) For every non-terminal node $u\in N$ and $b\in \set{0,1}$, $F(\child{u}{b})$ is the subfunction obtained from $F(u)$ by substituting $x_{\var(u)}$ with $b$, i.e.,
$F(\child{u}{b})=F(u)|_{x_{\var(u)}=b}$. 

\item
$\BDD{f}{\pi}$ must be minimal in the sense that
the following reduction rules cannot be applied.
In other words, $\BDD{f}{\pi}$ is obtained 
by maximally applying the following rules (\autoref{fig:rules}):
\begin{enumerate}
\item 
A node $u$ is \emph{redundant}
if $\child{u}{0}$ is the same node as $\child{u}{1}$.
If there exists a redundant node $u\in N$, then remove $u$ and its outgoing edges,
and redirect all the incoming edges of $u$ to $\child{u}{0}$.
\item
A pair of nodes $u$ and $v$ are \emph{equivalent}
if (1)  $\var(u)$ is equal to $\var(v)$, and
(2) $\child{u}{0}$ and $\child{u}{1}$ are the same nodes as $\child{v}{0}$ and $\child{v}{1}$, respectively.
If there exist a pair of {equivalent nodes}, $\set{u, v}\subset N$, then remove any one of them (say, $v$)
and its outgoing edges, and redirect all incoming edges of $v$ to $u$.
\end{enumerate}
\end{enumerate}
\end{definition}
\begin{remark}
Item 5 does not necessarily represent an actual way of constructing $\BDD{f}{\pi}$,
but just a way of defining $\BDD{f}{\pi}$.
For fixed $f$ and $\pi$, $\BDD{f}{\pi}$ is uniquely determined up to isomorphism regardless of the order of applying the rules in items 5.1 and 5.2~\cite{MeiThe98Book}.
\end{remark}
\begin{remark}\label{rem:def_mtbdd_zdd}
A zero-suppressed BDD (ZBDD, or ZDD)~\cite{Min93DAC}
is a variant of an OBDD. 
The only difference is the rule for redundant nodes.
Namely,
the definition of ZDDs
is obtained by replacing item~5.1 with 
"\emph{A node $u$ is \emph{redundant}
if $\child{u}{1}$ is the terminal node $\zero$.
If there exists a redundant node $u\in N$, then remove $u$ and its outgoing edges,
and redirect all the incoming edges of $u$ to $\child{u}{0}$}."

When the function $f$ has a multi-valued function:
$f\colon \set{0,1}^n\to [m] \subset \Natural$ for a fixed integer $m\ (\ge 3)$,
the corresponding variant of an OBDD is called
a multi-terminal BDD (MTBDD)~\cite{BahFroGaoHacMacParSom97FMSD,ClaMcmZhaFujYan97FMSD}.
The only difference from Boolean functions is that
each Boolean assignment is mapped
to an integer value, instead of a Boolean value.
Hence, the definition of MTBDD~\cite{BahFroGaoHacMacParSom97FMSD,ClaMcmZhaFujYan97FMSD} is obtained by just allowing the set of terminal nodes, $T$,
to accommodate $m$ nodes corresponding the $m$ values, e.g., $T=\set{\zero, \one_1, \one_2}$,
and modifying $F$ in item~4 accordingly.
\end{remark}

\begin{figure}[t] 
   \centering
   \includegraphics[width=13cm]{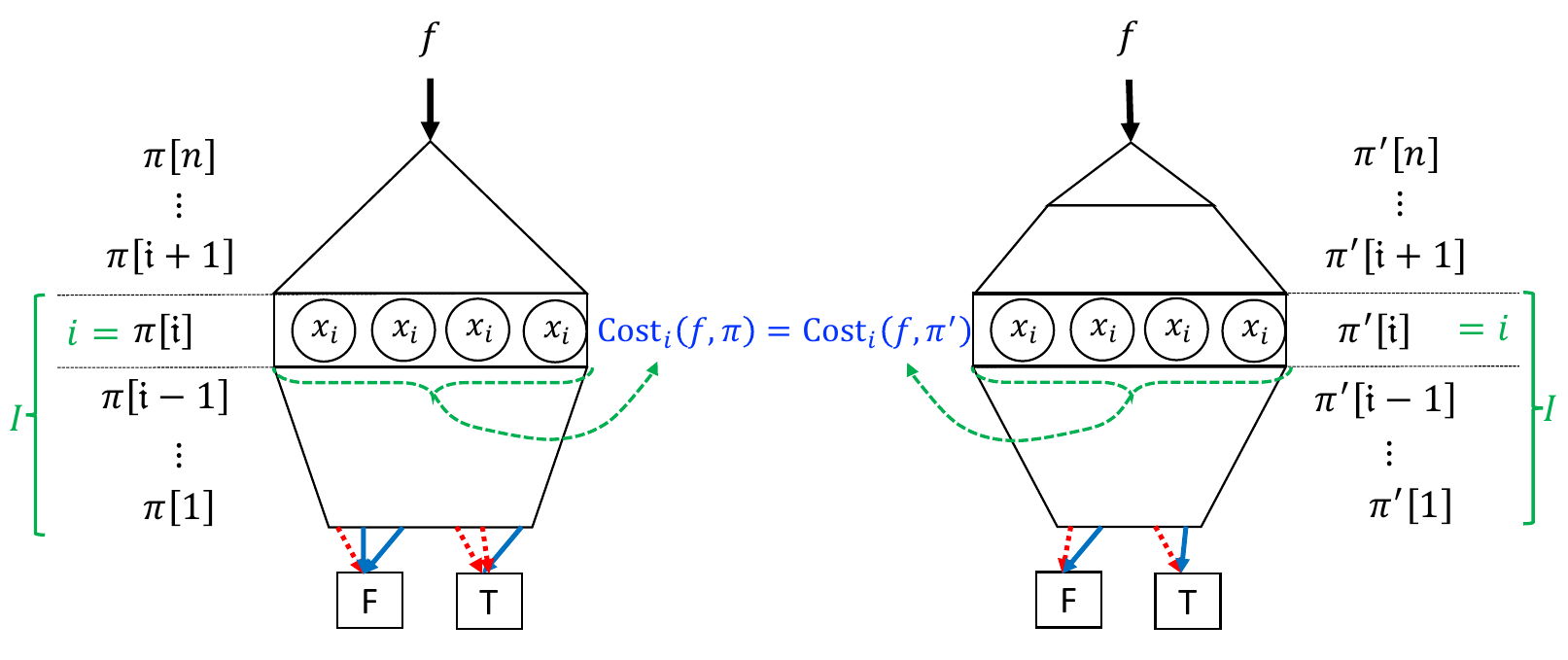} 
   \caption{Schematic representation of~\autoref{lm:OBDDwidth}:  For any subset $I\subseteq [n]$, suppose that two permutations
   $\pi,\pi' \in \calS_n$ satisfy $\set{\pi[1],\dots, \pi[\sizei-1]}=\set{\pi'[1],\dots, \pi'[\sizei-1]}$
   and $\pi[\sizei]=\pi'[\sizei]$, where $\sizei$ is the cardinaltiy of $I$. Then, it holds that the number of nodes labeled with $x_{i}$ in $\BDD{f}{\pi}$
   is equal to that  of nodes labeled with $x_{i}$ in $\BDD{f}{\pi'}$, where
   $i=\pi[\sizei]=\pi'[\sizei]$.}
   \label{fig:widthlemma}
\end{figure}

\begin{example}\label{example:OBDDs}
For ease of understanding the above notations, let us consider the OBDD 
on the right side
in \autoref{fig:OBDDs}.
The root  $\root$ is the uppermost node labeled with $x_6$.
The variable ordering $\pi$ is $(1,3,5,2,4,6)$. 
Every node in $V_\ell\ (\ell=1,\dots, 6)$ is represented by a circle labeled with $x_{\pi[\ell]}$.
For instance, $V_3$ consists of all the four nodes labeled with $x_5$.
For each node $v\in V_3$, it holds that $\var (v)=5$.
Let $u$ be the left node labeled with $x_3$.
Since the path from $\root$ to $u$ consists of three edges labeled with
$0$, $1$, and $0$ in this order from the root side,
$F(u)$ is represented as $F(\root)|_{x_6=0,x_4=1,x_2=0}=f|_{x_6=0,x_4=1,x_2=0}=x_3$.
\end{example}
\begin{figure}[t] 
   \centering
   \includegraphics[width=8cm]{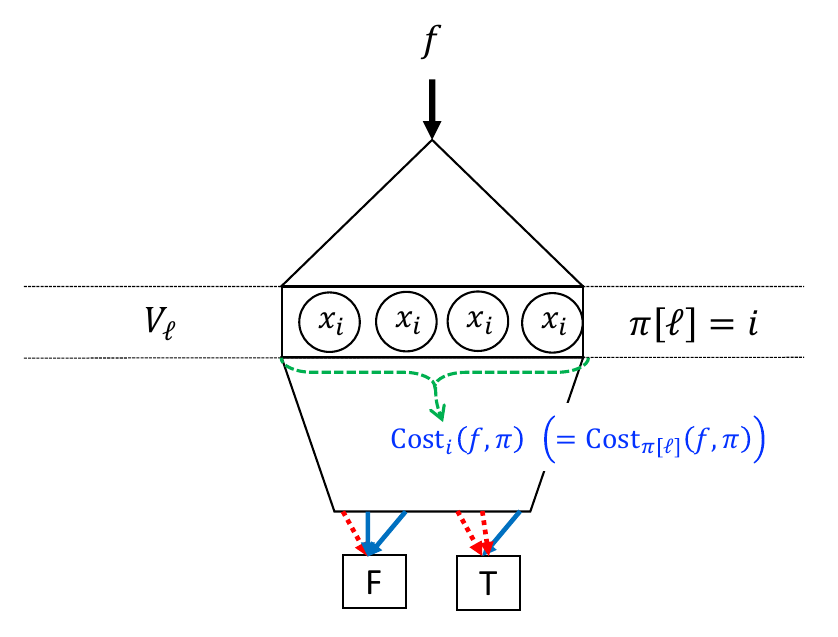} 
   \caption{Schematic representation of $\costfp{i}$}
   \label{fig:CostFunc}
\end{figure}
For each $i\in [n]$, 
$\costfp{i}$ denotes the width at the level associated with the variable $x_i$,
namely, the number of nodes
at the level $\pi^{-1}[i]$ (see \autoref{fig:CostFunc}).
For a subset $I\subseteq [n]$ of cardinality $\sizei$, let $\pi_I$ be a permutation $\pi$ in $\Pi(I)$
that minimizes the total number of nodes at level $1$ to level $\sizei$ over all $\pi\in \Pi(I)$:
\begin{equation}\label{eq:sumcost1}
\pi_I\deq \arg \min\Set{\sum_{\ell=1}^{\sizei} \costfp{\pi[\ell]}\colon {\pi\in \Pi(I)}}.
\footnote{We mean by \autoref{eq:sumcost1} and \autoref{eq:sumcost2} that  $\pi_I$ is one of the (possibly multiple) permutations that minimizes the sum in the right-hand side.}
\end{equation}
Note that $\sum_{\ell=1}^{\sizei} \costfp{\pi[\ell]}= \sum_{i\in I}\costfp{i}$
for $\pi\in \Pi(I)$.
More generally, 
for disjoint subsets $I_1,\dots, I_m\subseteq [n]$
of cardinality $\sizei_{1},\cdots,\sizei_{m}$, respectively,
$\pi_{\inner{I_1,\dots, I_m}}$ is a permutation in $\Pi(\inner{I_1,\dots, I_m})$
that minimizes the total number of
the nodes at level $1$ to level $\sizei_1+\dots +\sizei_m$
over all $\pi\in \Pi(\inner{I_1,\dots, I_m})$:
\begin{equation}\label{eq:sumcost2}
\pi_{\inner{I_1,\dots, I_m}}\deq \arg\min\Set{\sum_{\ell=1}^{\sizei_1+\dots +\sizei_m} \costfp{\pi[\ell]}\colon
\pi\in \Pi(\inner{I_1,\dots, I_m}) }.
\end{equation}
Note that 
$ {\sum_{\ell=1}^{\sizei_1+\dots +\sizei_m} \costfp{\pi[\ell]}}=\sum_{i\in I_1\sqcup \dots \sqcup I_m} \costfp{i}$
for any $\pi\in \Pi(\inner{I_1,\dots, I_m})$.
The following well-known lemma captures the essential property of OBDDs.
It states that
the number of nodes at level $\ell\in [n]$ is constant over all $\pi$ satisfying that the two sets $\set{\pi[1],\dots, \pi[\ell-1]}$ and $\set{\pi[\ell+1],\dots, \pi[n]}$ are fixed (see \autoref{fig:widthlemma}).

\begin{lemma}[\cite{FriSup90IEEETC}]\label{lm:OBDDwidth}
Let $f$ be a Boolean function over $n$ variables.
For any fixed non-empty subset $I\subseteq [n]$ of cardinality $\sizei$ and any fixed $i\in I$,
there exists a constant $c$ such that,
for  every $\pi\in \Pi(\inner{I\setminus \set{i},\set{i}})$,
$
\costfp{\pi[\sizei]}=\costfp{i}=c.
$
\end{lemma}

For convenience, we define  shorthand for the minimums of the sums in 
Eqs.~(\ref{eq:sumcost1}) and (\ref{eq:sumcost2}).
For $I'\subseteq I\subseteq [n]$,
$\mincost{I}[I']$ is defined as
the total number of nodes at the levels associated with variables indexed by elements in $I'$
under permutation  $\pi_I$, namely,
$
\mincost{I}[I']\deq \sum_{i\in I'}\cost{i}{f}{\pi_I}.
$
More generally, for disjoint subsets $I_1,\dots, I_m\subseteq [n]$ and $I'\subseteq I_1\sqcup\dots \sqcup I_m$,
\[
\mincost{\inner{I_1,\dots, I_m}}[I']
\deq
\sum_{i\in I'}
\cost{i}{f}{\pi_{\inner{I_1,\dots, I_m}}}.
\]
As a special case, $\mincost{\inner{I_1,\dots, I_m}}[I_1\sqcup\dots \sqcup I_m]$ is denoted by $\mincost{\inner{I_1,\dots, I_m}}$. We define $\mincost{\emptyset}$ as 0.

\subsection{The Algorithm by Friedman and Supowit}
This subsection reviews the algorithm by Friedman and Supowit~\cite{FriSup90IEEETC}.
We will generalize their idea later and heavily use the generalized form in our quantum algorithm.
Hereafter, we call their algorithm $\calg$.
\subsubsection{Key Lemma and Data Structures}
The following lemma is the basis of the dynamic programming approach used in $\calg$.
\begin{lemma}\label{lm:DPoriginal}
For any non-empty subset $I\subseteq [n]$ and any Boolean function $f\colon \set{0,1}^n\to \set{0,1}$,
the following holds:
$
\mincost{I}=\min_{i\in I}
\Oset{
\mincost{I\setminus i}+\cost{i}{f}{\pi_{\inner{I\setminus i,i}}}
}
=\min_{i\in I}
\Oset{
\mincost{\inner{I\setminus i,i}}
}.
$
\end{lemma}
\begin{proof}
Let $\sizei$ be the cardinality of $I$.
To show the first equality, let $i^*\deq \pi_I[\sizei]$. 
By definition, we have
\begin{align*}
\mincost{I}&=\sum_{i\in I}\cost{i}{f}{\pi_{I}}
= \sum_{i\in I\setminus{i^*}}\cost{i}{f}{\pi_{I}}+\cost{i^*}{f}{\pi_{I}}.
\end{align*}
The first term is equal to $\mincost{I\setminus i^*}$, since otherwise
there exists $\pi'\in \Pi(I)$ with  $\pi'[\sizei]=i^*$ and
$\pi'[\ell]\neq \pi_I[\ell]$ for some $\ell\in [\sizei-1]$
such that 
\begin{align*}
\sum_{i\in I}\cost{i}{f}{\pi'}&=
\sum_{i\in  I\setminus{i^*}}\cost{i}{f}{\pi'}+\cost{i^*}{f}{\pi'}\\
&< \sum_{i\in  I\setminus{i^*}}\cost{i}{f}{\pi_I}+\cost{i^*}{f}{\pi_I}
=\mincost{I},
\end{align*}
which contradicts the definition of $\mincost{I}$,
where we use $\cost{i^*}{f}{\pi'}=\cost{i^*}{f}{\pi_I}$ by \autoref{lm:OBDDwidth}.

The remaining term $\cost{i^*}{f}{\pi_{I}}$ is equal to
$\cost{i^*}{f}{\pi_{\inner{I\setminus i^*,i^*}}}$ by \autoref{lm:OBDDwidth}. 
Thus, the first equality in the statement of the lemma holds.
Since $\mincost{I\setminus i^*}=\sum_{i\in I\setminus i^*}\cost{i}{f}{\pi_{I\setminus i^*}}$
by the definition, and \autoref{lm:OBDDwidth} implies that $\cost{i}{f}{\pi_{I\setminus i^*}}=\cost{i}{f}{\pi_{\inner{I\setminus i^*,i^*}}}$ for every $i\in I\setminus i^*$,
it holds that $\mincost{I\setminus i^*}=\sum_{i\in I\setminus i^*}\cost{i}{f}{\pi_{\inner{I\setminus i^*,i^*}}}$.
Therefore, $\mincost{I\setminus i^*}+\cost{i^*}{f}{\pi_{\inner{I\setminus i^*,i^*}}}=\sum_{i\in I}\cost{i}{f}{\pi_{\inner{I\setminus i^*,i^*}}}$. This implies that the second equality in the lemma.
\end{proof}

Before sketching algorithm $\calg$,
we provide several definitions for any fixed Boolean function $f$ over $n$ variables.
For any subset $I\subseteq [n]$ of cardinality $\sizei$, 
$\tab_I$ is an array with $2^{n-\sizei}$ cells each of which stores a non-negative integer.
For each $\vb\in \set{0,1}^{n-\sizei}$,
the cell $\tab_I[\vb]$
stores (the pointer to) the unique node in $\BDD{f}{\pi_I}$ associated
via $F$ with function $f|_{x_{[n]\setminus I}=\vb}$.
Hence, we may write $\tab_I[x_{[n]\setminus I}=\vb]$ instead of $\tab_I[\vb]$
to clearly indicate the value assigned to each variable $x_\ell$ for $\ell\in [n]\setminus I$.
The purpose of $\tab_I$ is to relate
all subfunctions $f|_{\vx_{[n]\setminus I}=b}\ (\vb\in \set{0,1}^{n-\sizei})$ 
to the corresponding nodes in $\BDD{f}{\pi_I}$.
We assume without loss of generality that
the pointers to nodes in $\BDD{f}{\pi_I}$ are non-negative integers
and, in particular,  those to the two terminal nodes, $\zero$ and $\one$, corresponding to $\false$ and $\true$ are 
the integers 0 and 1, respectively.
Thus, $\tab_\emptyset$ is merely the truth table of $f$.

\begin{figure}[tb] 
   \centering
   \includegraphics[width=15cm]{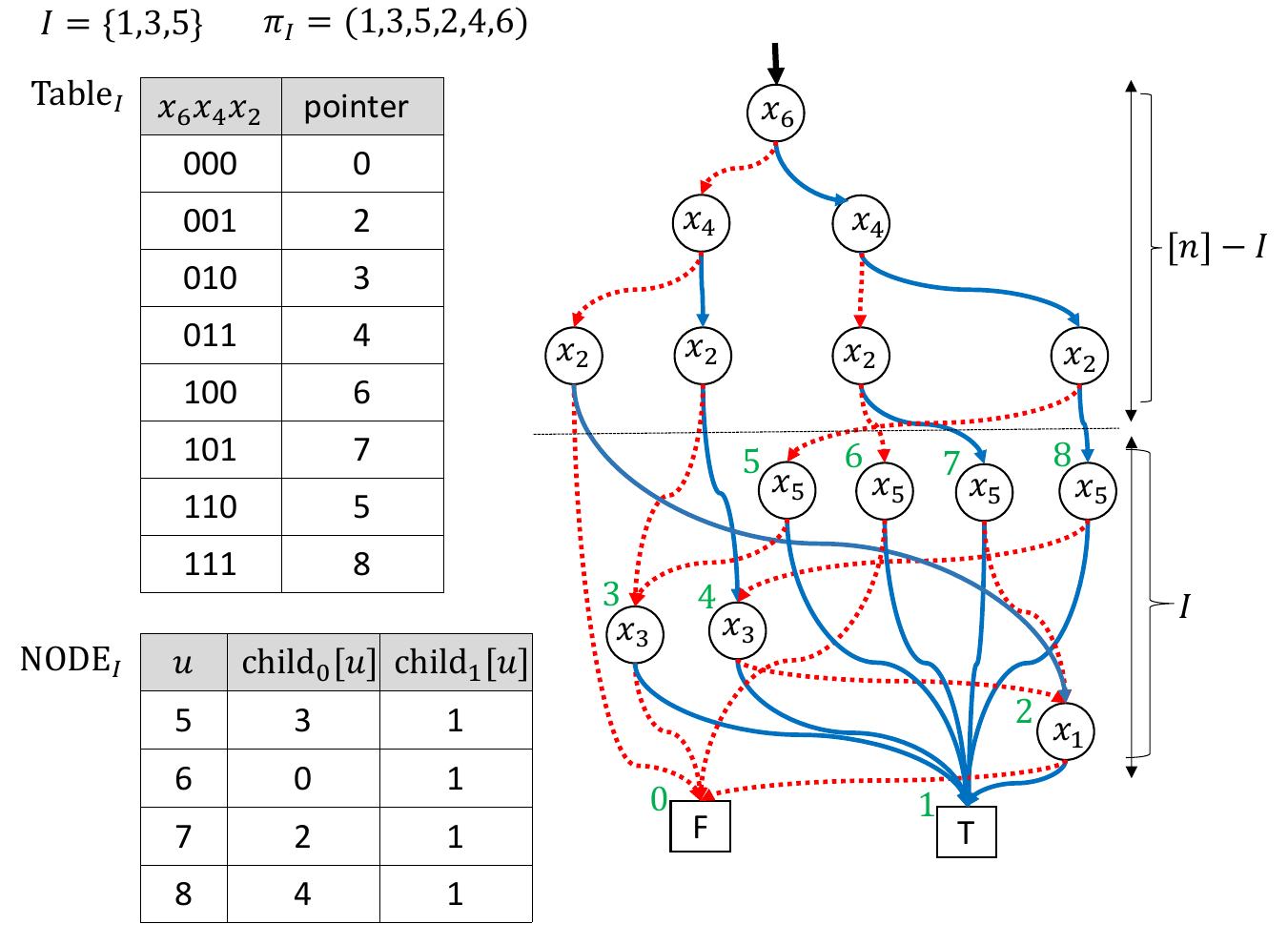} 
   \caption{Examples of data structures used in Algorithm $\calg$:
   $\tab_{I}$ and $\node_I$ with $I=\set{1,3,5}$ for the OBDD representing $f(x_1,\dots, x_{6})=x_1\cdot x_2+x_3\cdot x_4+\dots+x_{5}\cdot x_{6}$ for the variable ordering $(x_1,x_3, x_{5},x_2,x_4,x_{6})$. The pointers (integers) to the nodes labeled with $x_1, x_3, x_5$ are each shown at the top-left positions of the nodes.}
   \label{fig:SF}
\end{figure}

Algorithm $\calg$ computes $\tab_{I}$ 
together with 
$\pi_I$, $\mincost{I}$,
and another data structure, $\node_{I}$
for all subsets $I\subseteq[n]$,
starting from $\tab_{\emptyset}$ via dynamic programming.
Examples of $\tab_{I}$ and $\node_I$ are shown in \autoref{fig:SF}.
Intuitively, $\node_I$ stores
the subgraph of $\BDD{f}{\pi_I}$ induced by the outgoing edges of nodes in $V_{\sizei}$, where $\sizei$ is the cardinality of $I$.
More formally, $\node_I$ is the set of all triples
of (the pointers to) nodes,
$(u,\child{u}{0},\child{u}{1})$
for all $u\in V_{\sizei}$.
The purpose of the $\node_I$  is to prevent the algorithm
from duplicating existing nodes,
i.e., creating nodes associated with 
the same subfunctions
as those with which the existing nodes are associated.
By the definition, $\node_\emptyset$ is the empty set. We assume that $\node_I$ is implemented with an appropriate data structure, such as a balanced tree, so that the time complexity required for membership testing and insertion is the order of logarithm in the number of triples stored in $\node_I$.

More generally, for disjoint subset $I_1,\dots, I_m\subseteq [n]$
of cardinalities $\sizei_1,\dots, \sizei_m$, respectively,
$\tab_{\inner{I_1,\dots, I_m}}$ is an array with
$2^{n-(\sizei_1+\dots+ \sizei_m)}$ cells
such that, 
for $b\in \set{0,1}^{n-(\sizei_1+\dots+ \sizei_m)}$,
$\tab_{\inner{I_1,\dots, I_m}}[b]$
stores the nodes of $\BDD{f}{\pi_{\inner{I_1,\dots, I_m}}}$
associated with the function $f|_{x_{[n]\setminus I_1\sqcup\dots\sqcup I_m}=b}$.
$\node_{\inner{I_1,\dots, I_m}}$ is defined similarly for $\BDD{f}{\pi_{\inner{I_1,\dots, I_m}}}$.
For notational simplicity, we hereafter denote by $\FS{\inner{I_1,\dots, I_m}}$ the quadruplet 
\[
(\pi_{\inner{I_1,\dots, I_m}}, \mincost{\inner{I_1,\dots, I_m}}, \tab_{\inner{I_1,\dots, I_m}}, \node_{\inner{I_1,\dots, I_m}}).
\]

\subsubsection{Sketch of Algorithm $\calg$}
\begin{figure}[htbp] 
   \centering
   \includegraphics[height=8.cm]{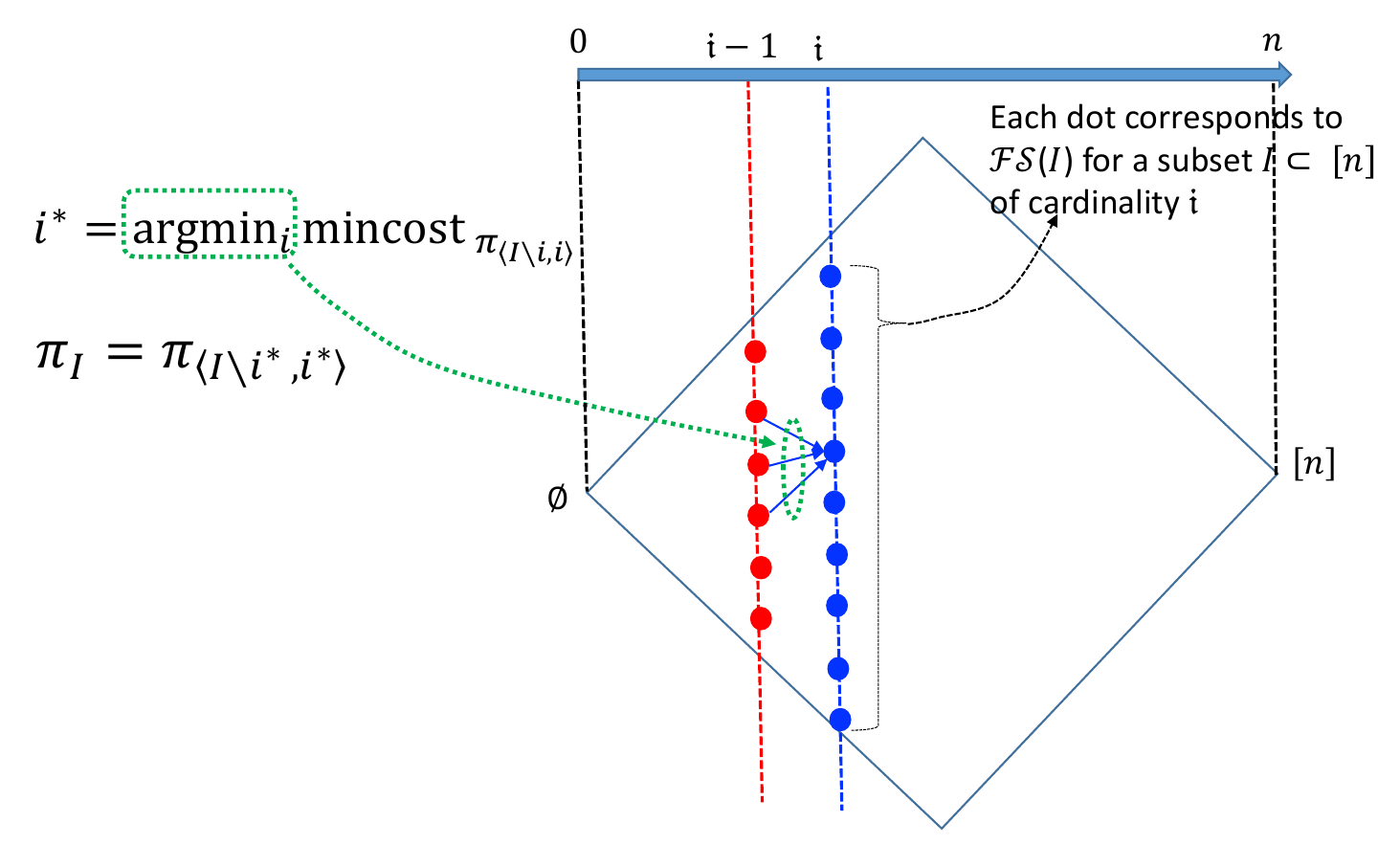} 
   \caption{Schematic view of Friedman-Supowit Algorithm.
   The algorithm goes from the left to the right.
   On the vertical line indicated by $\sizei$, there are $\binom{n}{\sizei}$ dots, each of which corresponds to 
   $\FS{I}$ for a subset $I\subseteq [n]$ of cardinality $\sizei$. $\FS{I}$ is computed from $\FS{\inner{I\setminus i}}$ for all $i\in I$, which are arranged as dots on the line indicated by $\sizei-1$ and   
have already been computed.}
   \label{fig:SchematicFS}
\end{figure}

Algorithm $\calg$ performs the following operations for $\sizei=1,\dots, n$ in this order.
For each subset $I\subseteq [n]$ of cardinality $\sizei$,
compute 
$\FS{\inner{I\setminus i,i}}$
from  $\FS{\inner{I\setminus i}}$
for each $i\in I$ in the manner described later 
(note that, since the cardinality of the set $I\setminus i$ is $\sizei-1$, 
$\FS{\inner{I\setminus i}}$ has already been computed).
Then set 
$
\FS{I}\longleftarrow\FS{\inner{I\setminus i^*,i^*}},
$
where $i^*$ is the index $i\in I$ that minimizes $\mincost{\inner{I\setminus i,i}}$,
meaning that $\pi_{I}$ is $\pi_{\inner{I\setminus i^*,i^*}}$.
This is justified by \autoref{lm:DPoriginal}.
A schematic view of the algorithm is shown in \autoref{fig:SchematicFS}.

To compute 
$\FS{\inner{I\setminus i,i}}$
from  $\FS{\inner{I\setminus i}}$, do the following.
First set $\node_{\inner{I\setminus i,i}}\leftarrow \emptyset$ 
and $\mincost{\inner{I\setminus i,i}}\leftarrow \mincost{{I\setminus i}}$ 
as their initial values.
Then, for each $\vb\in \set{0,1}^{n-\sizei}$, 
\begin{enumerate}
    \item Set 
    $u_0\leftarrow \tab_{I\setminus i}[x_{[n]\setminus I}=\vb,x_i=0]$ and 
$u_1\leftarrow \tab_{I\setminus i}[x_{[n]\setminus I}=\vb, x_i=1]$.
\item
If $u_0=u_1$, then register $u_0$ in $\tab_{\inner{I\setminus i,i}}[\vb]$%
\footnote{If one created a new node $u$ with $\child{u}{0}=u_0$ and $\child{u}{1}=u_1$, then the node $u$ would be redundant (see \autoref{fig:rules}).}
to associate $u_0$ with $f|_{x_{[n]\setminus I}=\vb}$.
Otherwise, test whether $(u,u_{0},{u}_{1})$ for some $u$ is stored in $\node_{\inner{I\setminus i,i}}$ in order not to create equivalent nodes.
If it is, 
the existing node $u$ is associated with $f|_{x_{[n]\setminus I}=\vb}$ and thus
register $u$ in the $\tab_{\inner{I\setminus i,i}}[\vb]$;
otherwise create a new triple $(u', {u}_{0},{u}_{1})$, insert it to $\node_{\inner{I\setminus i,i}}$
and increment $\mincost{\inner{I\setminus i,i}}$.
\end{enumerate}
Note that $u'$ must be different 
from any pointer already stored in $\node_{\inner{I\setminus i,i}}$
and
from any pointer to a node
in $V_1\sqcup \dots \sqcup V_{\sizei-1}$ 
in $\BDD{f}{\pi_{\inner{I\setminus i}}}$.
Such $u'$ can be easily chosen by setting $u'$ to
two plus the value of $\mincost{\inner{I\setminus i,i}}$ before the increment,
since the $\mincost{\inner{I\setminus i,i}}$ is exactly the number of triples in $\node_{\inner{I\setminus i,i}}$ plus $\abs{V_1\sqcup \dots \sqcup V_{\sizei-1}}$,
and the pointers 0 and 1 are reserved for the terminal nodes.
We call the above procedure \emph{table folding} with respect to $x_i$,
because it halves the size of $\tab_{\inner{I\setminus i}}$.
We also mean it by ``folding $\tab_{\inner{I\setminus i}}$ with respect to $x_i$''.

The complexity analysis is relatively simple.
For each $\sizei$, we need to compute 
$\FS{I}$
for $\binom{n}{\sizei}$ choices of subsets $I\subseteq [n]$ of cardinality $\sizei$.
For each such subset $I$, it takes $O^*(2^{n-\sizei})$ time 
since the the size of $\tab_{I\setminus i}$ is $2^{n-\sizei+1}$
and each operation to $\node_{I\setminus i}$ and $\node_{\inner{I\setminus i,i}}$ takes a polynomial time in $n$.
Thus, the total time is 
$
\sum_{\ell=0}^{n}2^{n-\ell+1}\binom{n}{\ell}=2\cdot 3^n
$
up to a polynomial factor. 
The point is that computing each 
$\FS{I}$
takes time linear to the size of $\tab_{I\setminus i}$ up to a polynomial factor.
The space required by Algorithm $\calg$ during the process for $\sizei$
is dominated by the sum of the space for $\tab_{I}$, $\tab_{I\setminus i}$ and $\node_{I}$ over all $I$ and $i\in I$,
which is $O^*\left(2^{n-\sizei}\binom{n}{\sizei}\right)$. 
The space complexity is thus $O^\ast\left(\max_{\ell\in \set{0}\cup[n]}2^{n-\ell}\binom{n}{\ell}\right)=O^\ast(3^n)$.

\begin{theorem}[Friedman and Supowit~\cite{FriSup90IEEETC}]\label{th:FS}
	Suppose that the truth table of $f\colon \set{0,1}^n \to \set{0,1}$ is given as input.
	Algorithm $\calg$ produces 
	$\FS{[n]}$
	in $O^\ast(3^n)$ time and space.
\end{theorem}

\subsection{Quantum Computation}
We assume that readers have a basic knowledge of quantum computing
(e.g., Refs.~\cite{NieChu00Book,KitSheVya02Book,KayLafMos07Book}).
We provide only a lemma used to obtain our results.

\begin{lemma}[Quantum Minimum 
Finding~\cite{DurHoy96ARXIV,BuhCleWolZal99FOCS,LeGMag18PODC}]
\label{lm:QuantumMinimumFinding}
For every $\varepsilon >0$ there exists a quantum algorithm 
that, for a function $f\colon [N]\to Z$ given as an oracle, where $Z$ is a finite totally ordered set,
finds an element $x\in [N]$ at which $f(x)$ achieves the minimum,
with error probability at most $\varepsilon$
by making $O(\sqrt{N\log (1/\varepsilon)})$ queries.
\end{lemma}

In this paper, the search space $N$ is exponentially large in $n$, and we are interested in 
exponential complexities, ignoring polynomial factors in them.
We can thus safely assume $\varepsilon=1/2^{p(n)}$ for a polynomial $p(n)$,
so that the overhead is polynomially bounded.
Since the depth
of recursions in our algorithms is at most constant
and each recursion level performs
the quantum minimum finding algorithm $2^{O(n)}$ times as a subroutine,
the total number of calling the subroutine
is $(2^{O(n)})^{O(1)}=2^{O(n)}$.
Thus,
the overall error probability is, by the union bound,
$2^{O(n)}/2^{p(n)}$,
which is exponentially small for a sufficiently large $p(n)$.
In the following proofs, we thus assume that $\varepsilon$ is exponentially small whenever we use 
\autoref{lm:QuantumMinimumFinding}.

Our algorithms assume the quantum random access memory (QRAM) model~\cite{GioLloMac08PRL},
which is commonly used in the literature when considering quantum algorithms.
QRAM is a quantum analog of random access memory (RAM). 
Recall that RAM allows you to read and write data at any position you specify.
QRAM provides this functionality \emph{in superposition}.
Suppose that QRAM stores classical data 
$D=(D_1,D_2, \cdots, D_n )$,
where $D_k\in \set{0,1}^m$ is the data stored at memory address $k\in [n]$.
The operation for reading out the data $D_k$ at address $k$ in QRAM
works as the following unitary operator:
\[
\ket{k}\otimes \ket{z}\otimes \ket{D}\mapsto \ket{k}\otimes \ket{z\oplus D_k}\otimes \ket{D}
\]
for every $z\in \set{0,1}^m$, where the third regsiter is QRAM.
By linearity, one can access $D_k$ in superposition over multiple $k$'s.
Similarly, QRAM allows you to write classical data $f(k)\in \set{0,1}^m$ to the location $k\in [n]$ in superposition:
\[
\ket{k}\otimes \ket{f(k)}\otimes \ket{z_1,\cdots,z_n}\mapsto \ket{k}\otimes \ket{f(k)}\otimes \Ket{z_1,\cdots, z_k\oplus f(k),\cdots,z_n},
\]
for every $z_i\in \set{0,1}^m$, where the third register is QRAM. 
The oracle used by quantum query algorithms models QRAM
(the oracle can also model the subroutine call to a quantum circuit, in which case
QRAM is not needed). In our case, we use QRAM to store the data computed in the preprocessing phase, as discussed in
\autoref{rm:QRAM} in  \autoref{subsec:SimpleCases}.

\section{Quantum Algorithm with Divide-and-Conquer}\label{sec:divide-and-conquer}
We generalize \autoref{lm:DPoriginal} and \autoref{th:FS} and use them in our quantum algorithm.
\begin{lemma}\label{lm:DPgeneral}
For any disjoint subsets $I_1,\dots,I_m,J\subseteq [n]$ with $J\neq \emptyset$ 
and any Boolean function $f\colon \set{0,1}^n\to \set{0,1}$,
the following holds:
\begin{align*}
\mincost{\inner{I_1,\dots,I_m,J}}
&=\min_{j\in J}
\Oset{
\mincost{\inner{I_1,\dots,I_m,J\setminus \set{j}}}+\cost{j}{f}{\pi_{\inner{I_1,\dots,I_m,J\setminus \set{j},\set{j}}}}
}\\
&=\min_{j\in J}
\Oset{
\mincost{\inner{I_1,\dots,I_m,J\setminus \set{j},\set{j}}}
}.
\end{align*}
\end{lemma}
The proof of this lemma is very similar to that of \autoref{lm:DPoriginal} and 
deferred to \autoref{appdx:ProofsOfLemmas}.
Based on \autoref{lm:DPgeneral}, we generalize \autoref{th:FS} to obtain algorithm~$\ccalg$
(its pseudo-code is given below, and
a schematic view of $\ccalg$ is shown in \autoref{fig:SchematicComposition}).
\begin{figure}[tbp] 
   \centering
   \includegraphics[height=7cm]{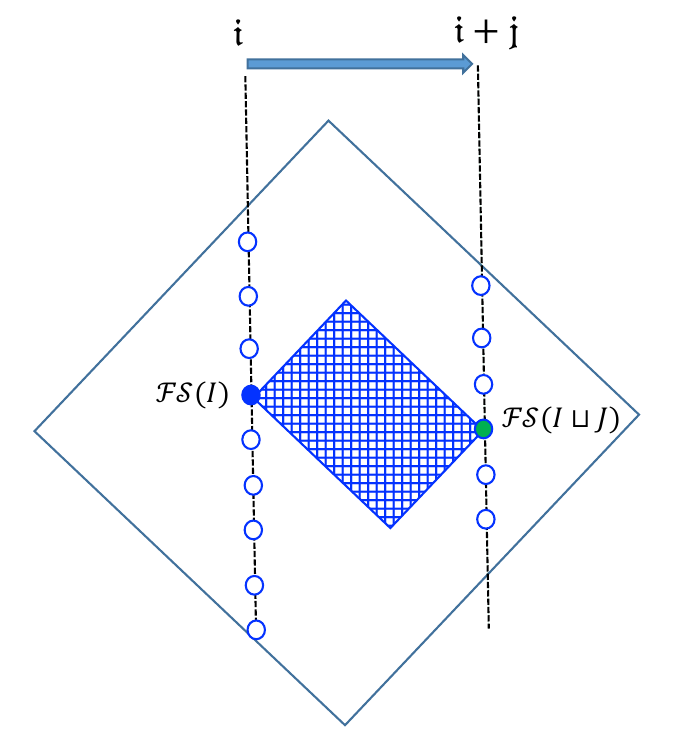} 
   \caption{Schematic view of $\ccalg$. This view corresponds to the case where
   $m=1$ and $J\subset [n]\setminus I$ in \autoref{lm:ClassicalComposition},
   where the cardinalities of $I$ and $J$ are $\sizei$ and $\sizej$, respectively. The shaded area is the one that $\ccalg$ sweeps to produce $\FS{\inner{I,J}}$.}
   \label{fig:SchematicComposition}
\end{figure}

\begin{lemma}[Classical Composition Lemma]\label{lm:ClassicalComposition}
For disjoint subsets $I_1,\dots, I_m, J\subseteq [n]$
of cardinalities $\sizei_1,\cdots, \sizei_m,\sizej$
with $J\neq \emptyset$,
there exists a deterministic algorithm~$\ccalg$ that
produces 
$\FS{\inner{I_1,\dots, I_m,J}}$
from $\FS{\inner{I_1,\dots, I_m}}$ for an underlying function $f\colon \set{0,1}^n \to \set{0,1}$
in $O^\ast\Oset{2^{n-(\sizei_1+\dots+\sizei_m+\sizej)}\cdot 3^{\sizej}}$ time and space.
More generally, 
for each $\sizek\in [\sizej]$,
the algorithm produces
the set $\set{\FS{\inner{I_1,\dots, I_m,K}}\colon K\subseteq J, \abs{K}=\sizek}$
from $\FS{\inner{I_1,\dots, I_m}}$
in $O^*\Oset{2^{n-(\sizei_1+\dots+\sizei_m+\sizej)}\sum_{\ell=0}^{\sizek}2^{\sizej-\ell}\binom{\sizej}{\ell}}$ 
time and space.
\end{lemma}
Note that if $I_1\sqcup\dots\sqcup I_m=\emptyset$ and $J=[n]$, then we obtain
 \autoref{th:FS}.

\begin{proof}
Let us focus on the simplest case of $m=1$,
for which our goal is to show an algorithm
that produces 
$\FS{\inner{I,J}}$
from 
$\FS{I}$.
Generalizing the proof to the case of $m\ge 2$ is straightforward.
Starting from $\FS{{I}}$, 
the algorithm first folds $\tab_{I}$ with respect to each variable in $\set{x_j\colon j\in J}$ 
to obtain $\FS{\inner{I,j}}$ for every $j\in J$.
Then,
for every $j_1,j_2\in J$,
it folds $\tab_{\inner{I,j_1}}$ with respect to $x_{j_2}$ and  $\tab_{\inner{I,j_2}}$ with respect to $x_{j_1}$ to
obtain $\FS{\inner{I,\set{j_1,j_2}}}$ 
by taking the minimum of $\mincost{\inner{I,j_1,j_2}}$
and $\mincost{\inner{I,j_2,j_1}}$;
it repeats this to finally
obtain $\FS{\inner{I,J}}$.
This algorithm is justified by~\autoref{lm:DPgeneral}.
For each $\sizek\in [\sizej]$, $K\subseteq J$ of cardinality $\sizek$, and $k\in K$,
the time complexity of computing
$\FS{\inner{I,K}}$
from 
$\FS{\inner{I,K\setminus k}}$
is linear to the size  of $\tab_{\inner{I,K\setminus k}}$, i.e.,  $2^{n-\sizei-\sizek+1}$ up to a polynomial factor.
The total time is thus, up to a polynomial factor,
\[
\sum_{\ell=1}^{\sizej}2^{n-\sizei-\ell}\binom{\sizej}{\ell}<
2^{n-\sizei-\sizej}\sum_{\ell=0}^{\sizej}2^{\sizej-\ell}\binom{\sizej}{\ell}
= 2^{n-(\sizei+\sizej)}\cdot 3^{\sizej}.
\]

If we stop the algorithm at $\ell=\sizek<\sizej$, then the algorithm produces
the set $\set{\FS{\inner{I,K}}\colon K\subseteq J, \abs{K}=\sizek}$. 
The time complexity in this case is at most
$2^{n-(\sizei+\sizej)}\sum_{\ell=0}^{\sizek}2^{\sizej-\ell}\binom{\sizej}{\ell}$
up to a polynomial factor.

Since the space complexity is trivially upper-bounded by the time complexity,
we complete the proof.
\end{proof}
\begin{remark}
One may think that the actual space complexity could be much less than the time complexity.
However, this is not the case.
The size of $\tab_{\inner{I,K}}$ is also the dominant factor determining the space complexity.
When computing $\tab_{\inner{I,K}}$ for a subset $K$ of cardinality $\sizek$,
it suffices to keep $\tab_{\inner{I,K}}$ and $\tab_{\inner{I,K\setminus k}}$ 
for every $k\in K$ in memory.
The space complexity is thus, up to a polynomial factor,
the maximum of 
$2^{n-\sizei-\sizek}\binom{\sizej}{\sizek}
+
2^{n-\sizei-(\sizek-1)}\binom{\sizej}{\sizek-1}$
over all $\sizek\in [\sizej]$, which is 
the same order as the time complexity.
\end{remark}

\begin{remark}
Note that one can prove lemmas
similar to Lemmas~\ref{lm:OBDDwidth}, \ref{lm:DPoriginal}, and \ref{lm:DPgeneral}
also in the case
of the OBDD variants, MTBDDs and ZDDs.
Thus,
Algorithm $\ccalg$ (almost) as is can output the optimal variable ordering and the MTBDD for that ordering, if the multi-value version of the truth table is given as input, since only
the difference from OBDDs is the number of terminal nodes (\autoref{rem:def_mtbdd_zdd}).
To apply $\calg$ to the case of ZDDs,
we need to modify the table folding rule to deal with redundant nodes. 
More concretely, we need to modify 
lines~\ref{ForZDD1} and \ref{ForZDD2} in \autoref{alg:cFS} follows:

\vspace{5pt}
{\indent \textbf{if} $u_1=0$ \textbf{then}\\
\indent\hspace{0.5cm}$\tab_{\inner{I, K-k,k}}[x_{[n]\setminus (I\sqcup K)}=\vb]\leftarrow u_0$}
\vspace{5pt}

\noindent These modifications are also possible for the quantum algorithms described later,
since they perform table folding by running~$\ccalg$ as a subroutine.
\end{remark}

\begin{algorithm}[tbh]
\LinesNumbered
\SetKw{DownTo}{downto}
\SetKwProg{Fn}{Function}{}{end}
\SetKwProg{Sr}{Subroutine}{}{end}
\SetKw{To}{to}
\KwIn{disjoint subsets $I, J\in [n]$ and $\FS{I}$, where $\sizei$ and $\sizej$ denote
the cardinalities of $I$ and $J$, respectively.}
\KwOut{$\FS{\inner{I,J}}$}
\BlankLine
\Fn{$\mathsf{Main}()$}{
\For{$\sizek:=1$ \To $\sizej$}{
\For{each $\sizek$-element subset $K\subseteq J$}{
$\mincost{\inner{I,K}}\leftarrow +\infty$\tcp*{init.}

\For{each $k\in K$}{
$\FS{\inner{I, K\setminus k,k}}\leftarrow \mathsf{FOLD}(I, K, k, \FS{\inner{I, K\setminus k}})$\;
\If{$\mincost{\inner{I,K}}> \mincost{\inner{I,K\setminus k,k}}$}
{
$\FS{\inner{I,K}}\leftarrow \FS{\inner{I, K\setminus k,k}}$\;
}
}
}
}
\KwRet{$\FS{\inner{I, J}}$}
}

\Fn(\tcp*[f]{produce $\FS{\inner{I, K\setminus {k},{k}}}$ from $\FS{\inner{I, K\setminus {k}}}$}
){$\mathsf{FOLD}(I, K, k, \FS{\inner{I, K\setminus k}})$}{
\BlankLine

$\mincost{\inner{I,K\setminus k,k}}\leftarrow \mincost{\inner{I,K\setminus k}}$\tcp*{init.}
$\node{\inner{I,K\setminus k,k}}\leftarrow \emptyset$\tcp*{init.}
\For(\tcp*[f]{$\sizek$ denotes $|K|$}){$\vb\in \set{0,1}^{n-\sizei-\sizek}$}{
$u_0\leftarrow \tab_{\inner{I, K\setminus k}}[x_{[n]\setminus(I\sqcup K)}=b,x_k=0]$;\

$u_1\leftarrow \tab_{\inner{I, K\setminus k}}[x_{[n]\setminus(I\sqcup K)}=b,x_k=1]$;\

\uIf{$u_0=u_1$\label{ForZDD1}}{$\tab_{\inner{I, K\setminus k,k}}[x_{[n]\setminus (I\sqcup K)}=\vb]\leftarrow u_0$\label{ForZDD2}}
\uElseIf{$\exists u\ (u,u_0,u_1)\in \node{\inner{I,K\setminus k,k}}$}{
$\tab_{\inner{I, K\setminus k,k}}[x_{[n]\setminus (I\sqcup K)}=\vb]\leftarrow u$
}
\Else(\tcp*[f]{create a new node}){
$u\leftarrow  \mincost{\inner{I,K\setminus k,k}}+2$\;
$\tab_{\inner{I, K\setminus k,k}}[x_{[n]\setminus(I\sqcup K)}=b]\leftarrow u$\;
$\mincost{\inner{I,K\setminus k,k}}\leftarrow \mincost{\inner{I,K\setminus k,k}}+1$\;
insert $(u,u_0,u_1)$ into $\node{\inner{I, K\setminus k,k}}$
}

}
set $\pi_{\inner{I, K\setminus k,k}}$ to an arbitrary $\pi \in \Pi(\inner{I, K\setminus k, k})$ such that $\pi[\ell]=\pi_{\inner{I, K\setminus k}}[\ell] \ (\ell=1,\dots, \sizei+\sizek-1)$;\

\KwRet{$\FS{\inner{I, K\setminus k,k}}$}
}
\SetAlgoRefName{{FS}$^*$}
\caption{Composable variant of algorithm $\calg$. ``$A\leftarrow B$'' means that $B$ is substituted for $A$.} \label{alg:cFS}
\end{algorithm}

\begin{figure}[tbp] 
   \centering
   \includegraphics[height=7cm]{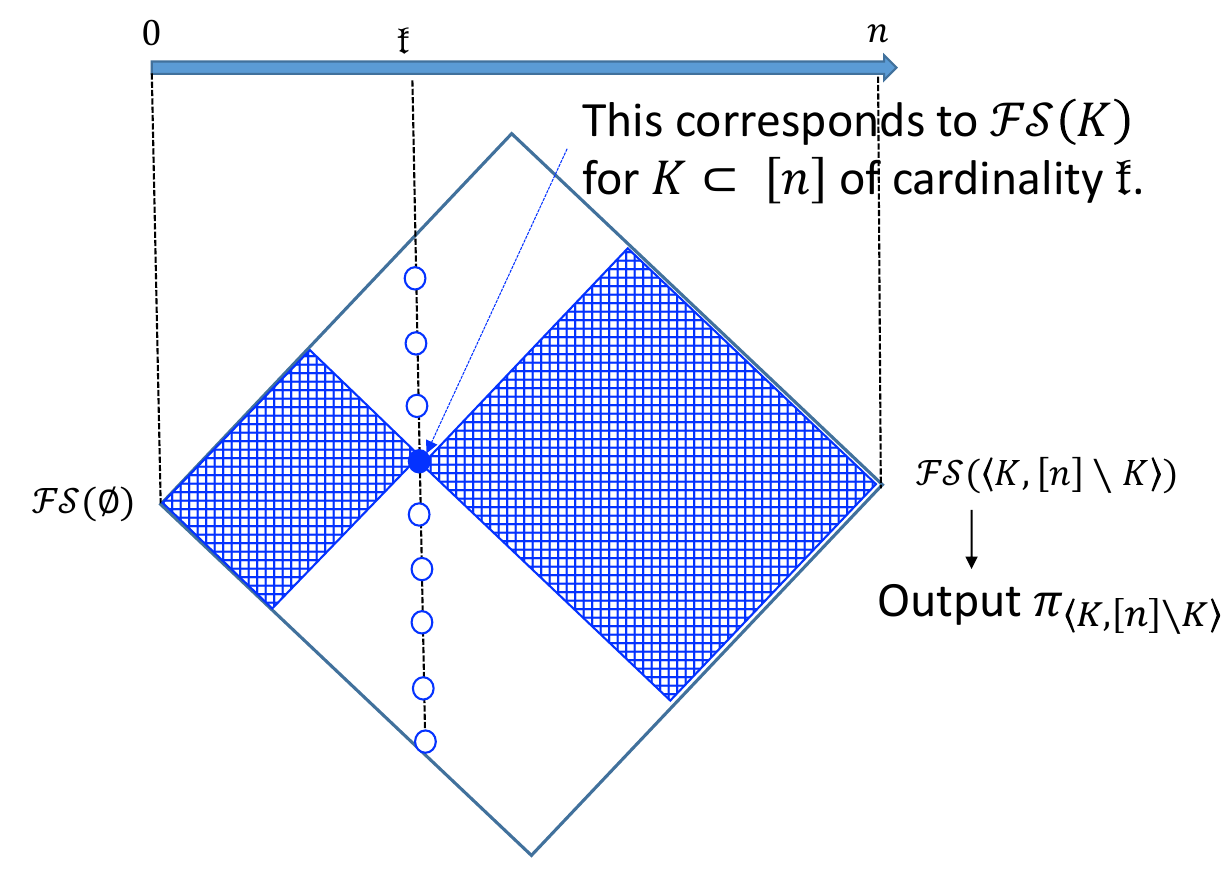} 
   \caption{Schematic view of Eq.~(\ref{eq:divide-and-conquer2}) in \autoref{lm:divide-and-conquer}. Intuitively, the lemma says that it is possible
   to decompose $\ccalg$ into the parts each of which 
   consists of the two shaded rectangles that share the dot corresponding to $\FS{K}$ on the line indicated by $\sizek$ 
   for a subset $K\subseteq [n]$ of fixed cardinality $\sizek$. The optimal variable ordering is induced by the decomposition for one of such $K$'s.}
   \label{fig:SchematicDivideConquer}
\end{figure}

The following lemma is the basis of our quantum algorithms.
\begin{lemma}[Divide-and-Conquer]\label{lm:divide-and-conquer}
For any disjoint subsets 
$I_1,\dots, I_m, J\subseteq [n]$ with $\sizej\deq\abs{J}\neq 0$
and any $\sizek\in [\sizej]$,
it holds that $\mincost{\inner{I_1,\dots, I_m,J}}[J]$
is equal to
\begin{equation}\label{eq:divide-and-conquer1}
\min_{K\colon K\subseteq J, \abs{K}=\sizek}
\Oset{
\mincost{\inner{I_1,\dots, I_m,K}}[K]
+\mincost{\inner{I_1,\dots, I_m,K,J\setminus K}}[J\setminus K]
}.
\end{equation}
In particular, when $I_1\sqcup\dots\sqcup I_m=\emptyset$ and $J=[n]$,
it holds that
\begin{equation}\label{eq:divide-and-conquer2}
\mincost{[n]}=
\min_{K\subseteq [n], \abs{K}=\sizek}
\Oset{
\mincost{K}
+\mincost{\inner{K,[n]\setminus K}}[[n]\setminus K]
}.
\end{equation}
\end{lemma}
A schematic view of the above lemma is shown in \autoref{fig:SchematicDivideConquer}.

\begin{proof}
 We first prove the special case of $I_1\sqcup\dots\sqcup I_m=\emptyset$ and $J=[n]$.
 By definition, we have 
\begin{align*}
\mincost{[n]} &= \sum_{\ell=1}^n \costfp{\pi[\ell]}= \sum_{\ell=1}^{\sizek} \costfp{\pi[\ell]} + \sum_{\ell=\sizek+1}^n \costfp{\pi[\ell]}
\end{align*}
for the optimal permutation $\pi\deq\pi_{[n]}$.
Let $K=\set{\pi[1],\dots,\pi[\sizek]}$.
By \autoref{lm:OBDDwidth}, the first sum is independent of how $\pi$ maps 
$\set{\sizek+1,\dots, n}$ to $[n]\setminus K$. 
Thus, it is equal to the minimum 
of $ \sum_{\ell=1}^{\sizek} \cost{\sigma[\ell]}{f}{\sigma}$
over all $\sigma\in \Pi(\inner{K,[n]\setminus K})$, i.e., $\mincost{K}$.
Similarly, the second sum is independent of how $\pi$ maps $[\sizek]$ to $K$.
Thus, it is equal to the minimum of $\sum_{\ell=\sizek+1}^n \cost{\sigma[\ell]}{f}{\sigma}$
over all $\sigma\in \Pi(\inner{K,[n]\setminus K})$, i.e., $\mincost{\inner{K,[n]\setminus K}}[[n]\setminus K]$.
This completes the proof of Eq.~(\ref{eq:divide-and-conquer2}).

We can straightforwardly generalize this. 
Let $\pi \deq \pi_{\inner{I_1,\dots, I_m,J}}$
and $\sizei\deq\abs{I_1\sqcup\dots \sqcup I_m}$.
Then, we have
\begin{align*}
\mincost{\inner{I_1,\dots, I_m,J}} [J]
&= \sum_{\ell=1}^{\sizek} \costfp{\pi[\sizei+\ell]} + \sum_{\ell=\sizek+1}^{\sizej} \costfp{\pi[\sizei+\ell]}.
\end{align*}
By defining
$K\deq \set{\pi[\sizei+1 ],\dots,\pi[\sizei+\sizek]}$,
the same argument as the special case of $\sizei=0$
implies that
the first and second sums are
$\mincost{\inner{I_1,\dots, I_m,K}}[K]$ and
$\mincost{\inner{I_1,\dots, I_m,K,J\setminus K}}[J\setminus K]$,
respectively. This completes the proof of Eq.~(\ref{eq:divide-and-conquer1}).
\end{proof}

\subsection{Simple Cases}\label{subsec:SimpleCases}
We provide simple quantum algorithms based on \autoref{lm:divide-and-conquer}.
The lemma states that, for any $\sizek\in [n]$,
$\mincost{[n]}$ is the minimum of 
$\mincost{K}
+\mincost{\inner{K,[n]\setminus K}}[[n]\setminus K]
$
over all subsets $K\subseteq [n]$ of cardinality $\sizek$.
To find $K$ 
from among $\binom{n}{\sizek}$ possibilities
that minimizes this amount,
we use the quantum minimum finding (\autoref{lm:QuantumMinimumFinding}).
To compute $\mincost{K}
+\mincost{\inner{K,[n]\setminus K}}[[n]\setminus K]=\mincost{\inner{K,[n]\setminus K}}$, 
it suffices to first compute $\FS{K}$ (including $\mincost{K}$), 
and then $\FS{\inner{K,[n]\setminus K}}$
(including $\mincost{\inner{K,[n]\setminus K}}$) from $\FS{K}$.
The time complexity for computing $\FS{K}$ from $\FS{\emptyset}$
is $O^\ast(2^{n-\sizek}3^{\sizek})$ by  \autoref{lm:ClassicalComposition}
with $I_1\sqcup \dots \sqcup I_m=\emptyset$ and $J=K$,
while that for computing $\FS{\inner{K,[n]\setminus K}}$
from $\FS{K}$
is $O^\ast(3^{n-\sizek})$ by  \autoref{lm:ClassicalComposition}
with $m=1$, $I_1\deq K$, and $J\deq [n]\setminus K$.
Thus, the time complexity for computing
$\FS{\inner{K,[n]\setminus K}}$ from $\FS{\emptyset}$
is $O^\ast(2^{n-\sizek}3^{\sizek}+ 3^{n-\sizek})$.
For $\sizek \deq\floor{\alpha n}$ with constant parameter $\alpha\in [0,1]$ fixed later,
the total time complexity up to a polynomial factor is
\[
T(n)=\sqrt{\binom{n}{\floor{\alpha n}}} \Oset{2^{(1-\alpha)n}3^{\alpha n}+3^{(1-\alpha)n}}
\lessapprox 2^{\frac{1}{2}\bentropy{\alpha}n}
\Set{
2^{\cset{(1-\alpha)+\alpha \log_2 3}n}
+
2^{\cset{(1-\alpha)\log_2 3}n}
}.
\]
To balance both terms, we set
$
(1-\alpha)+\alpha \log_2 3=(1-\alpha)\log_2 3
$
and obtain 
$\alpha=\alpha^\ast$, 
where
$
\alpha^\ast\deq \frac{\log_23-1}{2\log_23-1}\approx 0.269577.
$
We then have
\[
\min_{\alpha\in [0,1]}T(n)=O\Oset{2^{\frac{1}{2}\bentropy{\alpha^\ast}n+(1-\alpha^\ast)n+\alpha^\ast (\log_2 3)n}}=O(\gamma_0^n),
\]
where $\gamma_0\deq 2.98581\dots$. 
This slightly improves the classical best bound $O^\ast(3^n)$
on the time complexity.
To improve the bound further, we introduce a preprocessing phase
that classically computes 
$\FS{K}$ for every $K$ of cardinality $\floor{\alpha n} \ (\alpha \in (0,1))$
by using \autoref{alg:cFS}.
By \autoref{lm:ClassicalComposition},
the preprocessing time is then, up to a polynomial factor,
\begin{align}\label{eq:complexity_of_preprocess}
\sum_{\ell=1}^{\floor{\alpha  n}} 2^{n-\ell}\cdot \binom{n}{\ell}
&\le \alpha  n \cdot\max_{\ell\in [\floor{\alpha n}]} 2^{n-\ell}\binom{n}{\ell}
\lessapprox\left\{\begin{array}{cc}
2^{(1-\alpha )n+\bentropy{\alpha }n} & (\floor{\alpha n} \le n/3) \\
2^{\frac{2}{3}n+\bentropy{1/3}n} & (\floor{\alpha n}> n/3),
 \end{array}\right.
\end{align}
since $2^{n-\ell}\binom{n}{\ell}$ increases when $\ell\le n/3$
and decreases otherwise.
Note that once this preprocessing is completed,
we can use $\FS{K}$ for free
and assume that the cost for accessing the information in $\FS{K}$ is polynomially bounded
for all subsets $K\subseteq [n]$ of cardinality $\floor{\alpha n}$.

\begin{figure}[tbp] 
   \centering
   \includegraphics[height=7cm]{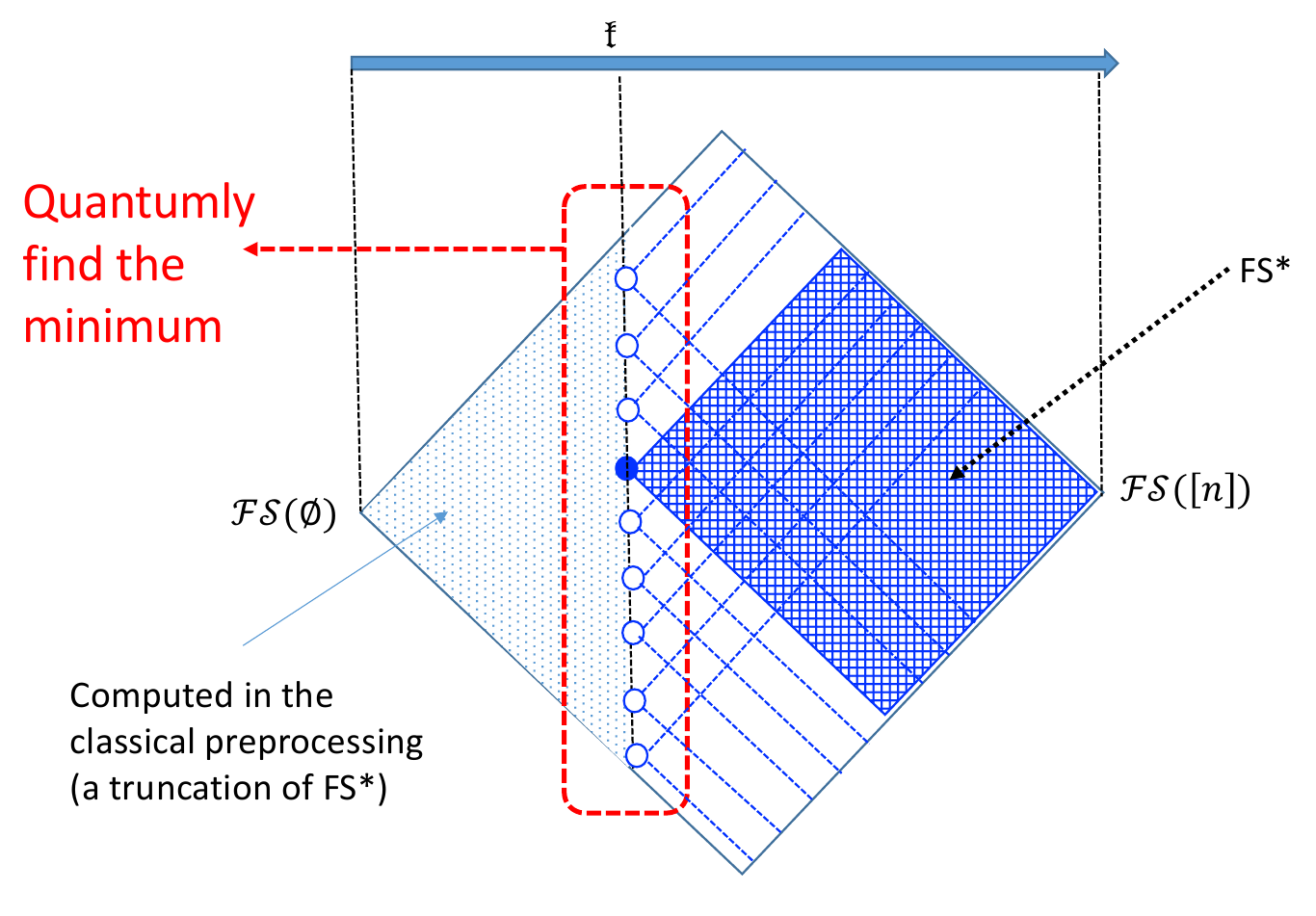} 
   \caption{Schematic view of our algorithm in the simplest case (one-parameter case). The dotted area
   is computed in the classical preprocessing, which is realized by truncating the process of $\ccalg$ as stated in \autoref{lm:ClassicalComposition}. Every shaded square area touching the line indexed by $\sizek$
      is computed by using $\ccalg$. The actual algorithm runs the quantum minimum finding, which calls $\ccalg$ to compute the shaded areas coherently.}
   \label{fig:Schematic1stAlg}
\end{figure}

Then, assuming that $\alpha<1/3$ (thus, $\floor{\alpha n}<n/3) $, the total time complexity up to a polynomial factor
is
\[
T(n)=\sum_{\ell=1}^{\floor{\alpha n}} 2^{n-\ell}\cdot \binom{n}{\ell}+\sqrt{\binom{n}{\floor{\alpha n}}}\Oset{
n^{O(1)}
+3^{(1-\alpha)n}}
\lessapprox
2^{\cset{(1-\alpha)+\bentropy{\alpha}}n}+2^{\cset{\frac{1}{2}\bentropy{\alpha}+(1-\alpha)\log_2 3}n}.
\]
To balance both terms, we set
$
(1-\alpha)+\bentropy{\alpha}=\frac{1}{2}\bentropy{\alpha}+(1-\alpha)\log_2 3$ and obtain the solution $\alpha = \alpha^\ast$, where $\alpha^\ast\deq 0.274863\dots$,
which is less than $1/3$, as we assumed.
At $\alpha=\alpha^\ast$, we have
$
T(n)\lessapprox 
2^{\cset{(1-\alpha^\ast)+\bentropy{\alpha^\ast}}n}=O^\ast(\gamma_1^n),
$
where $\gamma_1$ is at most $2.97625\ (<\gamma_0)$.
Thus,  introducing the preprocessing improves the complexity bound.
A schematic view of the above algorithm is shown in \autoref{fig:Schematic1stAlg}.
We provide the two-parameter case as a straightforward generalization of
the above single parameter case in \autoref{appdx:TwoParameters}.
We believe the two-parameter case is quite helpful to understand the general case in the following subsection.

\begin{remark}\label{rm:QRAM}
We use QRAM to store all the classical data $\FS{K}$ computed in the preprocessing phase
 (in the above and the following sections).
Even without QRAM, we can slightly improve the classical bound 
by computing $\FS{K}$  and then $\FS{\inner{K,[n]\setminus K}}$
over all $K$ in superposition.
With the help of QRAM, however, one can see that it is possible to achieve a better bound with preprocessing.
\end{remark}

\subsection{General Case}

We can improve this bound further by applying \autoref{lm:divide-and-conquer} $d$ times.
The resulting algorithm with constant parameters $d\in \Natural$ and $\vec{\alpha} \deq (\alpha_1,\dots,\alpha_d)$ is denoted by 
$\qalgorithm{d,\vec{\alpha}}$,
where $0<\alpha_1<\dots <\alpha_d<1$.
Its pseudo-code is given below.
In addition, we assume $\alpha_1<1/3$ and use a constant $\alpha_{d+1}\deq 1$ in the following complexity analysis.

\begin{algorithm}
\LinesNumbered
\SetKwProg{Fn}{Function}{}{end}
\SetKw{DownTo}{downto}
\KwIn{$\FS{\emptyset}:=\set{\tab_\emptyset, 
\pi_{\emptyset},
\mincost{\emptyset},
\node_{\emptyset}$\!\!}
 (accessible from all Functions)}
\KwOut{$\FS{[n]}$}
\BlankLine
\Fn{$\mathsf{Main}()$}{

compute the collection $\set{\FS{I}\colon I\subseteq [n], \ \abs{I}=\floor{\alpha_1n}}$
with algorithm $\calg$ (or $\ccalg$)\;\label{alg:plug-in1}
make the above collection global (i.e., accessible from all Functions)\;
\KwRet{$\mathsf{DivideAndConquer}([n],d+1)$}
}

\Fn(\tcp*[f]{Compute $\FS{J}$ with $\alpha_1,\dots, \alpha_r\  (\floor{\alpha_r n}=\abs{J})$}){$\mathsf{DivideAndConquer}(J,r)$}{
\lIf(\tcp*[f]{$\FS{J}$ has been precomputed.}){$r=1$}{
\KwRet{$\FS{J}$}%
}

find $K (\subset J)$ of cardinality $\floor{\alpha_{r-1}n}$,
with \autoref{lm:QuantumMinimumFinding},
that minimizes\label{alg:GroverHere} $\mincost{\inner{K,J\setminus K}}$,\\
\hspace{0.3cm} which is computed as a component of $\FS{\inner{K, J\setminus K}}$ 
by $\mathsf{ComputeFS}(K,J\setminus K,r)$\;
let $K^*$ be the set that achieves the minimum\;
\KwRet{$\FS{\inner{K^\ast,J\setminus K^\ast}}$}
}
\Fn(\tcp*[f]{Compute $\FS{\inner{K,H}}$ with $\alpha_1,\dots, \alpha_r\ (\floor{\alpha_r n}=\abs{K\sqcup H})$})
{$\mathsf{ComputeFS}(K,H,r)$}{
$\FS{K}\leftarrow \mathsf{DivideAndConquer}(K,r-1)$\;

$\FS{\inner{K,H}}\leftarrow \ccalg(K,H,\FS{K})$\;\label{alg:plug-in2}

\KwRet{$\FS{\inner{K,H}}$}
}

\SetAlgoRefName{OptOBDD$(d,{\alpha})$}
\caption{Quantum OBDD-minimization algorithm with constant parameters $d\in \Natural$ and $\vec{\alpha}\deq (\alpha_1,\dots, \alpha_d)\in [0,1]^d$ satisfying $0<\alpha_1<\dots< \alpha_d<1$,
where the quantum minimum finding algorithm is used in \autoref{alg:GroverHere}, and
$\ccalg$ (page~\pageref{alg:cFS}) is used in 
\autoref{alg:plug-in1} and  \autoref{alg:plug-in2}. ``$A\leftarrow B$'' means that $B$ is substituted for $A$.
} \label{alg:OptOBDD}
\end{algorithm}

To simplify notations, define two functions as follows:
for $x,y\in (0,1)$ such that $x<y$,
\begin{align*}
f(x,y)&\deq \frac{1}{2}y\cdot \Bentropy{{x/y}}  +g(x,y),\\
g(x,y)&\deq (1-y)+(y-x) \log_23.
\end{align*}

By \autoref{lm:ClassicalComposition},
the time required for the preprocessing
is $\sum_{\ell=1}^{\floor{\alpha_1 n}} 2^{n-\ell}\cdot \binom{n}{\ell}$
up to a polynomial factor.
Thus, the total time complexity can be described as the following recurrence:
\begin{align}
T(n)&=\sum_{\ell=1}^{\floor{\alpha_1 n}} 2^{n-\ell}\cdot \binom{n}{\ell}+L_{r+1}(n),\label{eq:T(n)}\\
L_{r+1}(n)&=\sqrt{\binom{\floor{\alpha_{r+1}n}}{\floor{\alpha_{r}n}}}\Oset{L_r(n)+2^{(1-\alpha_{r+1})n} 3^{(\alpha_{r+1}-\alpha_r)n}}\notag\\
&= \sqrt{\binom{\floor{\alpha_{r+1}n}}{\floor{\alpha_{r}n}}}\Oset{L_r(n)+2^{g(\alpha_r, \alpha_{r+1})n}},\label{eq:L_{j+1}}
\end{align}
where $r\in [d]$ and $L_1(n)=O^\ast(1)$.
Intuitively, $L_r(n)$ for $r\in [d]$ is the time required for producing
$\FS{\inner{K_r, K_r\setminus K_{r-1}}}$
for fixed $K_r$ of cardinality $\floor{\alpha_r n}$
such that $\mincost{\inner{K_{r-1}, K_r\setminus K_{r-1}}}$
is minimum over all subsets $K_{r-1}\ (\subset K_r)$
of cardinality $\floor{\alpha_{r-1}n}$
when the collection
of
$\FS{\inner{K_{r-1}, K_{r-1}\setminus K_{r-2}}}$
 is available at the cost of $L_{r-1}(n)$.

Since $L_1(n)=O^\ast(1)$, we have
\[
L_2(n)\lessapprox\sqrt{\binom{\floor{\alpha_2 n}}{\floor{\alpha_1 n}}}\cdot 2^{g(\alpha_1, \alpha_{2})n}
\lessapprox 2^{f(\alpha_1,\alpha_{2})n}.
\]
By setting $f(\alpha_1,\alpha_{2})=g(\alpha_2,\alpha_{3})$, we have
\[
L_3(n)
=\sqrt{\binom{\floor{\alpha_3 n}}{\floor{\alpha_2 n}}}\cdot (L_2(n)+2^{g(\alpha_2, \alpha_{3})n}) 
\lessapprox \sqrt{\binom{\floor{\alpha_3 n}}{\floor{\alpha_2 n}}}\cdot 2^{g(\alpha_2, \alpha_{3})n} 
\lessapprox 2^{f(\alpha_2,\alpha_{3})n}.
\]
In general, for $r=2,\dots,d$,
setting $f(\alpha_{r-1},\alpha_{r})=g(\alpha_r,\alpha_{r+1})$  yields
\[
L_{r+1}(n)\lessapprox 2^{f(\alpha_r,\alpha_{r+1})n}.
\]
Therefore, the total complexity \autoref{eq:T(n)} is
\[
T(n)\lessapprox {\sum_{\ell=1}^{\floor{\alpha_1 n}} 2^{n-\ell}\cdot \binom{n}{\ell}+2^{f(\alpha_d,\alpha_{d+1})n}}
\lessapprox {2^{(1-\alpha_1)n+\bentropy{\alpha_1}n}+2^{f(\alpha_d,1)n}},
\]
where we use 
$\alpha_1<1/3$, $\alpha_{d+1}=1$, and 
\autoref{eq:complexity_of_preprocess}.
To optimize the right-hand side, we set parameters so that $1-\alpha_1+\bentropy{\alpha_1}=f(\alpha_d,1)$.

In summary, we need to find the values of parameters $\alpha_1,\dots, \alpha_d$ that satisfy the following system of equations and $\alpha_1<1/3$:
\begin{align}
1-\alpha_1+\bentropy{\alpha_1}&=f(\alpha_d,1),\label{eq:optimize_alpha1}\\
f(\alpha_{r-1},\alpha_{r})&=g(\alpha_r,\alpha_{r+1})\label{eq:optimize_alpha2}\ \ \ \ (r=2,\dots, d).
\end{align}
By numerically solving this system of equations, we 
obtain $T(n)=O(\gamma_d ^n)$, 
where $\gamma_d$ is  at most $2.83728$ for $d=6$.
The value of $\gamma_d$ becomes smaller as $d$ increases. However, incrementing $d$ beyond 6 provides only negligible improvement of $\gamma_d$. 
We show the numerical data in \autoref{tab:data_divideandconquer} in \autoref{appdx:NumericalData}.

Lastly, the number of calling the quantum minimum finding is at most $(2^n)^6=2^{6n}$. The total error probability is $1/2^{n^2-6n}$ when we set $\varepsilon = 1/2^{n^2}$ in \autoref{lm:QuantumMinimumFinding}.
Since the space complexity is trivially upper-bounded
by the time complexity, we have the following theorem.
\begin{theorem}\label{th:quantum_algorithm_preprocess}
There exists a quantum algorithm that,
for the truth table of $f\colon \set{0,1}^n\to\set{0,1}$ given as input,
produces $\FS{[n]}$ 
with probability $1-\exp (-\Omega(n^2))$
in $O^\ast(\gamma^n)$ time and space, where the constant $\gamma$ is
at most $2.83728$, which is achieved by 
$\qalgorithm{k,\vec{\alpha}}$
with $k=6$ and 
\sloppy
$
\vec{\alpha}=( 0.183791, 0.183802, 0.183974, 0.186131, 0.206480, 0.343573).$
\end{theorem}
Note that the values of $\alpha_i$'s are not symmetric with respect to $1/2$. 
This reflects the fact that optimizing cost is not symmetric  with respect to $1/2$,
contrasting with many other combinatorial problems.

\section{Quantum Algorithm with Composition}\label{sec:composition}
\subsection{Quantum Composition Lemma}
By generalizing
the quantum algorithm given in \autoref{th:quantum_algorithm_preprocess},
we now provide a quantum version of \autoref{lm:ClassicalComposition},
called the \emph{quantum composition lemma}.

\begin{lemma}[Quantum Composition: Base Part]\label{lm:QuantumComposition:BaseCase}
For any disjoint subsets $I_1,\dots, I_m,   J\subseteq [n]$
of cardinalities $\sizei_1, \dots, \sizei_m, \sizej\neq 0$, respectively, 
there exists a quantum algorithm
that, with probability $1-\exp(-\Omega(n^2))$,
produces 
$\FS{\inner{I_1,\dots, I_m,J}}$
from $\FS{\inner{I_1,\dots, I_m}}$ for an underlying function $f\colon \set{0,1}^n \to \set{0,1}$
in $O^\ast\Oset{2^{n-(\sizei_1+\dots +\sizei_m+\sizej)}\cdot \gamma^{\sizej}}$ time
and space,
where the constant $\gamma$ is 
at most $2.83728$
(the constant defined in \autoref{th:quantum_algorithm_preprocess}).
\end{lemma}
A pseudo-code of the algorithm provided in \autoref{lm:QuantumComposition:BaseCase}
is shown as $\cqalgorithm{\Gamma}{d,\vec{\alpha}}$ on page~\pageref{alg:cOptOBDD},
where 
the subroutine $\Gamma$
appearing in \autoref{alg2:plug-in}
is set to the deterministic algorithm~$\ccalg$,
and
$d$ and $\vec{\alpha}$ are set to the values specified 
in \autoref{th:quantum_algorithm_preprocess}.

\begin{proof}
The proof idea is similar to that used in the proof of \autoref{lm:ClassicalComposition}.

Since the space complexity is trivially upper-bounded by the time complexity,
we only analyze the time complexity in the following.
For simplicity, we assume $m=1$, write just $I$ instead of $I_1$, and let $\sizei$ be the cardinality of $I$.
It is straightforward to generalize to the case of $m\ge 2$.

We now provide the algorithm $\cqalgorithm{\Gamma}{d,\vec{\alpha}}$ on page \pageref{alg:cOptOBDD}
that produces 
$\FS{\inner{I,J}}$
from $\FS{\inner{I}}$,
where the subroutine $\Gamma$ used in \autoref{alg2:plug-in} is set to algorithm~$\ccalg$.
As parameters, the algorithm has an integer $d\in \Natural$ and a vector $\vec{\alpha}\deq (\alpha_1,\dots,\alpha_d)\in (0,1)^d$
such that $0<\alpha_1<\dots<\alpha_d<1$. Set $d$ and $\vec{\alpha}$ to the same values assumed in
the algorithm in \autoref{th:quantum_algorithm_preprocess}.
In addition, we use the constant $\alpha_{d+1}=1$ in the following.

Let $\sizej$ be the cardinality of ${J}$. 
In the preprocessing, the algorithm 
computes 
the collection $\set{\FS{\inner{I,K}}\colon K\subseteq J, \abs{K}=\floor{\alpha_1\cdot\sizej}}$ 
based on \autoref{lm:ClassicalComposition}
for given $\FS{I}$ with
the time complexity
$2^{n-\sizei-\sizej} \sum_{\ell=1}^{\floor{\alpha_1\cdot \sizej}} 2^{\sizej-\ell}\binom{\sizej}{\ell}$
up to a polynomial factor.
Thus, the total time complexity is expressed as
\begin{align}\label{eq:T'(n,n')}
T'(n,\sizej)&=2^{n-\sizei-\sizej} \sum_{\ell=1}^{\floor{\alpha_1 \sizej}} 2^{\sizej-\ell}\binom{\sizej}{\ell}+L'_{d+1}(n,\sizej),
\end{align}
where $L'_{d+1}(n,\sizej)$ is the time taken to perform all but the preprocessing.

Based on \autoref{lm:divide-and-conquer},
the algorithm proceeds in a way similar to the one given in \autoref{th:quantum_algorithm_preprocess},
which corresponds to the special case of $I\deq\emptyset$ and $J\deq[n]$.
The complexity $L'_{d+1}(n,\sizej)$ is then  expressed by the following recurrence, up to polynomial factors:
\begin{align*}
L'_{r+1}(n,\sizej)&=\sqrt{\binom{\floor{\alpha_{r+1}\cdot\sizej}}{\floor{\alpha_{r}\sizej}}}\Oset{L'_r(n,\sizej)+2^{n-\sizei-\alpha_{r+1}\cdot\sizej} 3^{(\alpha_{r+1}-\alpha_r)\cdot \sizej}}\ \ \ \ \ [r\in [d]],\\
L'_1(n,\sizej)&=O^\ast(1).
\end{align*}

In the following, we prove by induction that $T'(n,\sizej)\lessapprox 2^{n-\sizei-\sizej}T(\sizej)$,
where the function $T(\cdot)$ is defined in \autoref{eq:T(n)}.
Since 
$T(\sizej)=O^{*}(\gamma^{\sizej})$, this completes the proof for $m=1$.

Since $L'_1(n,\sizej)=O^\ast(1)$,  we have
\begin{align*}
L'_2(n,\sizej)&\lessapprox 2^{n-\sizei-\sizej} \sqrt{\binom{\floor{\alpha_2\cdot \sizej}}{\floor{\alpha_1\cdot \sizej}}}\ 
2^{(1-\alpha_2)\cdot\sizej}3^{(\alpha_2-\alpha_1)\cdot\sizej}=2^{n-\sizei-\sizej} L_{2}(\sizej),
\end{align*}
where the function $L_{2}(\cdot)$ is defined in \autoref{eq:L_{j+1}} for $r=1$.
This is the base case of the induction.
Then, assuming that $L'_{r}(n,\sizej)\lessapprox 2^{n-\sizei-\sizej}L_{r}(\sizej)$, we have
\begin{align*}
L'_{r+1}(n,\sizej)&\lessapprox \sqrt{\binom{\floor{\alpha_{r+1}\cdot\sizej}}{\floor{\alpha_{r}\cdot\sizej}}}\Oset{2^{n-\sizei-\sizej}L_{r}(\sizej)+2^{n-\sizei-\alpha_{r+1}\cdot\sizej} 3^{(\alpha_{r+1}-\alpha_r)\cdot\sizej}}\\
&=2^{n-\sizei-\sizej} \sqrt{\binom{\floor{\alpha_{r+1}\cdot\sizej}}{\floor{\alpha_{r}\cdot\sizej}}}\Oset{L_{r}(\sizej)+2^{\sizej-\alpha_{r+1}\cdot\sizej} 3^{(\alpha_{r+1}-\alpha_r)\cdot\sizej}}=2^{n-\sizei-\sizej}L_{r+1}(\sizej).
\end{align*}
Therefore, it holds that $L'_{d+1}(n,\sizej)\lessapprox 2^{n-\sizei-\sizej}L_{d+1}(\sizej)$ by induction.
Then, it follows from 
\autoref{eq:T'(n,n')} that
\begin{align*}
T'(n,\sizej)&\lessapprox 2^{n-\sizei-\sizej} \sum_{\ell=1}^{\floor{\alpha_1\cdot \sizej}} 2^{\sizej-\ell}\binom{\sizej}{\ell}+2^{n-\sizei-\sizej}L_{d+1}(\sizej)
=2^{n-\sizei-\sizej}T(\sizej).
\end{align*}
Since $T(\sizej)=O^\ast(\gamma^{\sizej})$ by \autoref{th:quantum_algorithm_preprocess},
it holds that 
$T'(n,\sizej)=O^\ast(2^{n-(\sizei+\sizej)}\gamma^{\sizej})$.
The total error probability is 
$(2^n)^d \cdot \varepsilon$ by the union bound, which is $\exp(-\Omega(n^2))$
by setting $\varepsilon=1/2^{n^2}$
in \autoref{lm:QuantumMinimumFinding}.
\end{proof}

\begin{algorithm}[tbh]
\LinesNumbered
\SetKwProg{Fn}{Function}{}{end}
\KwIn{disjoint subsets $I,J\subseteq [n]$, where $\sizej=\abs{J}$, and $\FS{I}$. (accessible from all Functions)}
\KwOut{$\FS{\inner{I,J}}$}
\BlankLine
\Fn{$\mathsf{Main}()$}{

compute the set $\set{\FS{\inner{I,K_1}}\colon K_1\subseteq J, \ \abs{K_1}=\floor{\alpha_1\cdot \sizej}}$
with algorithm $\ccalg$\;
make $\sizej$ and the above set of $\FS{\inner{I,K_1}}$ global (i.e., accessible from all Functions)\;
\KwRet{$\mathsf{DivideAndConquer}(I,J,d+1)$}
}

\Fn(\tcp*[f]{Compute $\FS{\inner{I,J}}$ with $\alpha_1,\dots, \alpha_r$})
{$\mathsf{DivideAndConquer}(I,J,r)$}{
\lIf(\tcp*[f]{$\FS{I,J}$ has been precomputed.}){$r=1$}{
\KwRet{$\FS{I,J}$}%
}
find $K (\subset J)$ of cardinality $\floor{\alpha_{r-1}\cdot\sizej}$,
with \autoref{lm:QuantumMinimumFinding},
that minimizes\label{alg2:GroverHere} $\mincost{\inner{I,K,J\setminus K}}$\\
\hspace{0.3cm}which is computed as a component of 
$\FS{\inner{I,K,J\setminus K}}$ \\
\hspace{0.3cm}by calling $\mathsf{ComputeFS}(I,K,J\setminus K,r)$\;
let $K^*$ be the set that achieves the minimum\;
\KwRet{$\FS{\inner{I, K^\ast,J\setminus K^\ast}}$}
}
\Fn(\tcp*[f]{Compute $\FS{\inner{I,K,H}}$ with $\alpha_1,\dots, \alpha_r $})
{$\mathsf{ComputeFS}(I,K,H,r)$}{
$\FS{I, K}\leftarrow \mathsf{DivideAndConquer}(I,K,r-1)$\;
$\FS{\inner{I, K,H}}\leftarrow \Gamma (I,K,H, \FS {I,K})$\;\label{alg2:plug-in}
\KwRet{$\FS{\inner{I,K,H}}$}
}
\SetAlgoRefName{OptOBDD$^\ast_{\Gamma}(d, \alpha)$}
\caption{Composable Quantum OBDD-minimization algorithm with subroutine $\Gamma$ and constant parameters $d\in \Natural$ and 
$\vec{\alpha}=(\alpha_1,\dots, \alpha_d)\in [0,1]^d$ satisfying $0<\alpha_1<\dots< \alpha_d<1$,
where the quantum minimum finding algorithm is used in \autoref{alg2:GroverHere},
and subroutine $\Gamma$ is used in \autoref{alg2:plug-in}.
``$A\leftarrow B$'' means that $B$ is substituted for $A$.
$\Gamma (I,K,H, \FS{I,K})$ is supposed to produce $\FS{\inner{I,K,H}}$ from $\FS{\inner{I,K}}$.
}\label{alg:cOptOBDD}
\end{algorithm}

\begin{lemma}[Quantum Composition: Induction Part]\label{lm:QuantumComposition:InductionPart}
Suppose that $\Gamma$ is a quantum algorithm that,
for any disjoint subsets $I_1,\dots, I_m, J\subseteq [n]$ 
of cardinalities $\sizei_1,\dots, \sizei_m,\sizej\neq 0$,
produces 
$\FS{\inner{I_1,\dots, I_m,J}}$ from $\FS{\inner{I_1,\dots, I_m}}$ 
with probability at least $1-\exp(-\Omega(n^2))$
in $O^\ast\Oset{2^{n-(\sizei_1+\dots+ \sizei_m+ \sizej)}\cdot \gamma^{\sizej}}$ time and space
for an underlying function $f\colon \set{0,1}^n \to \set{0,1}$.
Then, for any constant parameters $d \in \Natural$ and $\vec{\alpha}=(\alpha_1,\dots, \alpha_d)\in (0,1)^d$ 
with $\alpha_1<\dots<\alpha_d$
and for any disjoint subsets $I_1,\dots, I_m, J\subseteq [n]$ 
of cardinalities $\sizei_1,\dots, \sizei_m,\sizej\neq 0$,
$\cqalgorithm{\Gamma}{d,\vec{\alpha}}$
produces $\FS{\inner{I_1,\dots, I_m,J}}$ from $\FS{\inner{I_1,\dots, I_m}}$
with probability $1-\exp(-\Omega(n^2))$
in $O^\ast\Oset{2^{n-(\sizei_1+\dots+ \sizei_m+ \sizej)}\cdot \beta_d^{\sizej}}$ time and space
for the function $f$,
where $\beta_d^n$ upper-bounds, up to a polynomial factor, the time 
required for 
$\cqalgorithm{\Gamma}{d,\alpha}$ to 
compute $\FS{[n]}$ from $\FS{\emptyset}$,
that is, $T(n)=O^\ast(\beta_d^n)$,
where $T(n)$ is the function satisfying the following recurrence:
\begin{align}
T(n)&=\sum_{\ell=1}^{\floor{\alpha_1 n}} 2^{n-\ell} \binom{n}{\ell}+L_{d+1},\label{eq:QC:T(n)}\\
L_{r+1}&=\sqrt{\binom{\floor{\alpha_{r+1}n}}{\floor{\alpha_{r}n}}}\Oset{L_r+2^{(1-\alpha_{r+1})n} \gamma^{(\alpha_{r+1}-\alpha_r)n}}
= \sqrt{\binom{\floor{\alpha_{r+1}n}}{\floor{\alpha_{r}n}}}\Oset{L_r+2^{g_\gamma(\alpha_r, \alpha{r+1})}}, \label{eq:QC:L_{j+1}}
\end{align}
where $r\in [d]$, $L_1=O^\ast(1)$ and 
$g_\gamma (x,y)\deq (1-y)+(y-x) \log_2\gamma$.
\end{lemma}
\begin{proof}
Recall that 
algorithm $\ccalg$
is used as a subroutine in $\qalgorithm{d,\vec{\alpha}}$
provided in \autoref{th:quantum_algorithm_preprocess}.
Since the input and output of $\Gamma$ assumed in the statement are the same as
those of algorithm~$\ccalg$,
one can use $\Gamma$ instead of algorithm~$\ccalg$
in $\qalgorithm{d,\vec{\alpha}}$
(compromising on an exponentially small error probability).
Let $\qalg _{\Gamma}(d,\vec{\alpha})$ be the resulting algorithm.
Then, one can see that the time complexity $T(n)$ of 
$\qalg _{\Gamma}(d,\vec{\alpha})$
satisfies the recurrence:
Eqs.~(\ref{eq:QC:T(n)})-(\ref{eq:QC:L_{j+1}}),
which are obtained by just replacing $g(x,y)$ with $g_\gamma(x,y)$
in 
Eqs.~(\ref{eq:T(n)})-(\ref{eq:L_{j+1}}).
Suppose that $T(n)=O^\ast(\beta_d^n)$ follows from the recurrence.

Next, we generalize $\qalg _{\Gamma}(d,\vec{\alpha})$
so that it produces $\FS{\inner{I_1,\dots, I_m,J}}$ from $\FS{\inner{I_1,\dots, I_m}}$
for any disjoint subsets $I_1,\dots, I_m, J\subseteq [n]$ 
of cardinalities $\sizei_1,\dots, \sizei_m,\sizej\neq0$.
The proof is similar to that of \autoref{lm:QuantumComposition:BaseCase}.
The only difference is that
the time complexity of $\Gamma$ is 
$O^\ast\Oset{2^{n-(\sizei_1+\dots+ \sizei_m+ \sizej)}\cdot \gamma^{\sizej}}$, instead of 
$O^\ast\Oset{2^{n-(\sizei_1+\dots+ \sizei_m+ \sizej)}\cdot 3^{\sizej}}$.
Namely, when $m=1$ and $\sizej=\abs{J}$, the time complexity of $\cqalgorithm{\Gamma}{d,\vec{\alpha}}$
satisfies the following recurrence, up to polynomial factors:
\begin{align*}
T'(n,\sizej)&=2^{n-\sizei-\sizej} \sum_{\ell=1}^{\floor{\alpha_1 \cdot\sizej}} 2^{\sizej-\ell}\binom{\sizej}{\ell}+L'_{d+1}(n,\sizej),\\
L'_{r+1}(n,\sizej)&=\sqrt{\binom{\floor{\alpha_{r+1}\cdot\sizej}}{\floor{\alpha_{r}\cdot\sizej}}}\Oset{L'_r(n,\sizej)+2^{n-\sizei-\alpha_{r+1}\cdot\sizej} \gamma^{(\alpha_{r+1}-\alpha_r)\cdot\sizej}}\ \ \ \ \ [r\in [d]],\\
L'_1(n,\sizej)&=O^\ast(1),
\end{align*}
from which it follows that
$T'(n,\sizej)=2^{n-\sizei-\sizej}\cdot T(\sizej)=O^\ast\Oset{2^{n-(\sizei+\sizej)}\cdot \beta_d^{\sizej}}$.
It is straightforward to generalize to the case of $m\ge 2$.

Since the number of calling $\Gamma$ plus the number of using the quantum minimum finding outside $\Gamma$
is at most 
$(2^{n})^d+(2^{n})^d=2^{dn+1}$,
the total error probability
$2^{dn+1}\cdot (\exp(-\Omega(n^2))+\varepsilon)=\exp(-\Omega(n^2))$, where $\varepsilon = 1/2^{n^2}$ is the error probability
of the quantum minimum finding outside $\Gamma$.
\end{proof}

\subsection{The Final Algorithm}
Lemmas~\ref{lm:QuantumComposition:BaseCase} and \ref{lm:QuantumComposition:InductionPart}
naturally lead to the following algorithm.
We first define $\Gamma^{(1)}$ as $\cqalgorithm{\ccalg}{d^{(0)},\vec{\alpha}^{(0)}}$ for some $d^{(0)}\in \Natural$ and $\vec{\alpha}^{(0)}\in [0,1]^{d^{(0)}}$.
Then, we define $\Gamma^{(2)} $ as $\cqalgorithm{\Gamma_1}{d^{(1)},\vec{\alpha}^{(1)}}$ for some $d^{(1)}\in \Natural$ and $\vec{\alpha}^{(1)}\in [0,1]^{d^{(1)}}$.
In this way, we can define $\Gamma^{(\ell+1)}$ as $\cqalgorithm{\Gamma^{(\ell)}}{d^{(\ell)},\vec{\alpha}^{(\ell)}}$ for some $d^{(\ell)}\in \Natural$ and $\vec{\alpha}^{(\ell)}\in [0,1]^{d^{(\ell)}}$.
Accordingly, let $\gamma^{(\ell)}$ and $\beta^{(\ell)}_{d^{(\ell)}}$ be $\gamma$ and $\beta_d$ for each $\ell$.

Fix $d^{(\ell)}=6$ for every $\ell$.
Note that, in the proofs of 
Lemmas~\ref{lm:QuantumComposition:BaseCase} and \ref{lm:QuantumComposition:InductionPart},
parameter $\vec{\alpha}^{(\ell)}=(\alpha^{(\ell)}_1,\dots, \alpha^{(\ell)}_6)\in [0,1]^6$
 is set for each $\ell$
so that
it satisfies the system of equations,
straightforward generalizations of Eqs.~(\ref{eq:optimize_alpha1})-(\ref{eq:optimize_alpha2}),
\begin{align}
1-\alpha^{(\ell)}_1+\bentropy{\alpha^{(\ell)}_1}&=f_\gamma(\alpha^{(\ell)}_6,1),\\
f_\gamma(\alpha^{(\ell)}_{r-1},\alpha^{(\ell)}_{r})&=g_\gamma(\alpha^{(\ell)}_r,\alpha^{(\ell)}_{r+1}) \hspace{1cm}(r=2,\dots, 6),
\end{align}
where
$f_\gamma(x,y) \deq \frac{1}{2}y\cdot \Bentropy{{x}/{y}}  +g_\gamma(x,y)$ and
$g_\gamma (x,y)\deq (1-y)+(y-x) \log_2\gamma$.

By numerically solving this system of equations for $\ell=0$,
where $\gamma^{(0)}=3$, we have $\beta^{(0)}_6<2.83728$
as shown in \autoref{th:quantum_algorithm_preprocess}.
Then, numerically solving the system of equations for $\ell=1$
with $\gamma^{(1)}\deq 2.83728>\beta^{(0)}_6$, we have $\beta^{(1)}_6<2.79364$.
In this way, we obtain a certain $\beta_6^{(9)}$ less than $2.77286$ at the tenth composition.
We show the numerical data in \autoref{tab:data_composition} in \autoref{appdx:NumericalData}.
We therefore obtain the following theorem.
\begin{theorem}\label{th:quantum_algorithm_composition}
There exists 
a quantum algorithm that,
for the truth table of $f\colon \set{0,1}^n\to\set{0,1}$ given as input,
produces $\FS{[n]}$ in $O^\ast(\gamma^n)$ time and space
with probability $1-\exp(-\Omega(n^2))$, where the constant $\gamma$ is
at most $2.77286$.
\end{theorem}

\section{Conclusion}\label{sec:conclusion}
We have provided a quantum algorithm that solves the optimal variable ordering problem, one of the central problems concerning OBDDs, 
in $O^\ast(\gamma^n)$ time and space in the quantum random access memory (QRAM) model, where constant $\gamma$ is at most 2.77286 and  $n$ is the number of variables on which the input Boolean function depends.
This implies an exponential speedup over the best known classical algorithm, which runs in $O^\ast(3^n)$ time and space.

There are several questions that we have left open. 
The first question is whether improving the time and/or space complexity is possible.
We do not believe that our complexity bound is (nearly) tight. 
Our algorithm has an exponentially small error probability, 
with which its output, i.e., a variable ordering $\pi$ and the OBDD according to $\pi$, is not optimal. However, the output OBDD is still a valid OBDD for input function $f$, that is, neither an OBDD not representing $f$ nor a bit string not representing any OBDD.
It would be interesting to consider the case of
allowing the output that is not a valid OBDD with a small probability.

Second, a trivial classical lower bound is $\Omega(2^n)$, since the input is the truth table of a Boolean function over $n$ variables. 
This also holds even in the quantum setting, since the quantum query complexity
of identifying all $N$ bits hidden in an input oracle is $\Omega(N)$ in the worst case~\cite{AmbIwaNakNisRayTanYam09ARXIV,Kot14STACS}.
Is it possible to provide a better lower bound (under some conjectures)?

Third, we have considered time-efficient algorithms.
Their space complexity is equal to their time complexity, up to polynomial factors,
since the dominant part of each algorithm is the preprocessing
(because the preprocessing cost and the rest are made balanced), which has the same order of time and space complexities.
However, space complexity would sometimes be more critical than time complexity.
Hence, another exciting direction is the quest for space-efficient algorithms.
Actually, the $O^\ast(2^n)$ space complexity can be achieved
by the brute-force algorithm,
which, for {every} permutation $\pi\in \Pi([n])$,
constructs $\BDDfp$  in $O^\ast(2^n)$ time 
by restricting the $\calg$-algorithm to the fixed variable ordering $\pi$.
However, this incurs the enormous time complexity of $O^\ast(n!2^n)$.
In general, there is a trade-off between time and space.
Thus, a reasonable direction would be to make the product $TS$ of time complexity $T$ and space complexity $S$ as small as possible.
For instance, the product $TS$ of the brute-force algorithm is $O^\ast(n!4^n)$,
while the best known classical algorithm $\calg$ has $TS=O^\ast(3^n\cdot 3^n)=O^\ast(9^n)$.
Our algorithm has a much smaller $TS$ value of $O^\ast(2.77286^{2n})=O^\ast(7.68875^{n})$.
Is it possible to provide an algorithm that achieves a lower $TS$ value? 

Finally, the divide-and-conquer lemma (\autoref{lm:divide-and-conquer}) does not depend on quantum computing. It would be interesting to find classical applications of this lemma.

\section*{Acknowledgements}
This work is partially supported by JSPS KAKENHI Grant Number JP20H05966 and JP22H00522, and JST [Moonshot R\&D – MILLENNIA Program] Grant No. JPMJMS2061.
\bibliographystyle{plain}


\begin{thebibliography}{10}

\bibitem{AmbBalIraKokPruVih19SODA}
Andris Ambainis, Kaspars Balodis, J\={a}nis Iraids, Martins Kokainis,
  Kri\v{s}j\={a}nis Pr\={u}sis, and Jevg\={e}nijs Vihrovs.
\newblock Quantum speedups for exponential-time dynamic programming algorithms.
\newblock In {\em Proceedings of the Thirtieth Annual {ACM-SIAM} Symposium on
  Discrete Algorithms, {SODA} 2019}, pages 1783--1793, 2019.

\bibitem{AmbIwaNakNisRayTanYam09ARXIV}
Andris Ambainis, Kazuo Iwama, Masaki Nakanishi, Harumichi Nishimura, Rudy
  Raymond, Seiichiro Tani, and Shigeru Yamashita.
\newblock Average/worst-case gap of quantum query complexities by on-set size.
\newblock {\em arXiv}, 0908.2468, 2009.

\bibitem{AroBar09Book}
Sajeev Arora and Boaz Barak.
\newblock {\em Computational Complexity: A Modern Approach}.
\newblock Cambridge University Press, 2009.

\bibitem{BahFroGaoHacMacParSom97FMSD}
R.~I. Bahar, E.~A. Frohm, C.~M. Gaona, G.~D. Hachtel, E.~Macii, A.~Pardo, and
  F.~Somenzi.
\newblock Algebric decision diagrams and their applications.
\newblock {\em Formal Methods in System Design}, 10(2--3):171--206, 1997.

\bibitem{BolWeg96IEEETC}
B.~{Bollig} and I.~{Wegener}.
\newblock Improving the variable ordering of {OBDDs} is {NP}-complete.
\newblock {\em IEEE Transactions on Computers}, 45(9):993--1002, Sep. 1996.

\bibitem{Bol16ToCS}
Beate Bollig.
\newblock On the minimization of (complete) ordered binary decision diagrams.
\newblock {\em Theory of Computing Systems}, 59(3):532--559, Oct 2016.

\bibitem{Bry86IEEETC}
Randal~E. Bryant.
\newblock Graph-based algorithms for boolean function manipulation.
\newblock {\em IEEE Trans. Comput.}, 35(8):677--691, August 1986.

\bibitem{Bry91IEEETC}
Randal~E. {Bryant}.
\newblock On the complexity of {VLSI} implementations and graph representations
  of boolean functions with application to integer multiplication.
\newblock {\em IEEE Transactions on Computers}, 40(2):205--213, Feb 1991.

\bibitem{Bry92CSUR}
Randal~E. Bryant.
\newblock Symbolic boolean manipulation with ordered binary-decision diagrams.
\newblock {\em ACM Comput. Surv.}, 24(3):293--318, September 1992.

\bibitem{Bry18BookChapter}
Randal~E. Bryant.
\newblock Binary decision diagrams.
\newblock {\em Handbook of Model Checking}, pages 191--217, 2018.

\bibitem{BryChe95DAC}
Randal~E. Bryant and Yirng{-}An Chen.
\newblock Verification of arithmetic circuits with binary moment diagrams.
\newblock In {\em Proceedings of the 32nd Conference on Design Automation},
  pages 535--541, 1995.

\bibitem{BuhCleWolZal99FOCS}
Harry~M. Buhrman, Richard~E. Cleve, Ronald de~Wolf, and Christof Zalka.
\newblock Bounds for small-error and zero-error quantum algorithms.
\newblock In {\em Proceedings of the 40th Annual Symposium on Foundations of
  Computer Science, {FOCS} '99}, pages 358--368, 1999.

\bibitem{ClaMcmZhaFujYan97FMSD}
E.M. Clarke, K.~L. Mcmillan, X.~Zhao, M.~Fujita, and J.~Yang.
\newblock Spectral transforms for large boolean functions with applications to
  technology mapping.
\newblock {\em Formal Methods in System Design}, 10(2--3):137--148, 1997.

\bibitem{DreBec98Book}
Rolf Drechsler and Bernd Becker.
\newblock {\em Binary Decision Diagrams: Theory and Implementation}.
\newblock Springer, 1998.

\bibitem{DurHeiHoyMha06SICOMP}
Christoph D{\"u}rr, Mark Heiligman, Peter H{\o}yer, and Mehdi Mhalla.
\newblock Quantum query complexity of some graph problems.
\newblock {\em SIAM Journal on Computing}, 35(6):1310--1328, 2006.

\bibitem{DurHoy96ARXIV}
Christoph D\"{u}rr and Peter H{\o}yer.
\newblock A quantum algorithm for finding the minimum.
\newblock Technical Report quant-ph/9607014, arXiv, 1996.

\bibitem{FriSup90IEEETC}
S.~J. {Friedman} and K.~J. {Supowit}.
\newblock Finding the optimal variable ordering for binary decision diagrams.
\newblock {\em IEEE Transactions on Computers}, 39(5):710--713, May 1990.

\bibitem{GarJoh79Book}
Michael~R. Garey and David~S. Johnson.
\newblock {\em {COMPUTERS AND INTRACTABILITY} --- {A} Guide to the Theory of
  {NP}-Completeness}.
\newblock {W. H. Freeman and Company}, New York, 2 edition, 1979.

\bibitem{GioLloMac08PRL}
Vittorio Giovannetti, Seth Lloyd, and Lorenzo Maccone.
\newblock Quantum random access memory.
\newblock {\em Phys. Rev. Lett.}, 100:160501, Apr 2008.

\bibitem{Gro96STOC}
Lov~K. Grover.
\newblock A fast quantum mechanical algorithm for database search.
\newblock In {\em Proceedings of the 28th Annual {ACM} Symposium on Theory of
  Computing}, pages 212--219, 1996.

\bibitem{HeaMer94IEEETC}
M.~A. {Heap} and M.~R. {Mercer}.
\newblock Least upper bounds on {OBDD} sizes.
\newblock {\em IEEE Transactions on Computers}, 43(6):764--767, June 1994.

\bibitem{Hea93JET}
Mark Heap.
\newblock On the exact ordered binary decision diagram size of totally
  symmetric functions.
\newblock {\em Journal of Electronic Testing}, 4(2):191--195, 1993.

\bibitem{HehChe92IEEETC}
{Heh-Tyan Liaw} and {Chen-Shang Lin}.
\newblock On the {OBDD}-representation of general boolean functions.
\newblock {\em IEEE Transactions on Computers}, 41(6):661--664, June 1992.

\bibitem{HeiMol00IEEETC}
L.~{Heinrich-Litan} and P.~{Molitor}.
\newblock Least upper bounds for the size of {OBDDs} using symmetry properties.
\newblock {\em IEEE Transactions on Computers}, 49(4):360--368, April 2000.

\bibitem{HorIba02AI}
Takashi Horiyama and Toshihide Ibaraki.
\newblock Ordered binary decision diagrams as knowledge-bases.
\newblock {\em Artif. Intell.}, 136(2):189--213, 2002.

\bibitem{HorYaj97ISAAC}
Takashi Horiyama and Shuzo Yajima.
\newblock Exponential lower bounds on the size of {OBDDs} representing integer
  divistion.
\newblock In {\em Proceedings of the 8th International Symposium on Algorithms
  and Computation, ({ISAAC} '97)}, Lecture Notes in Computer Science, pages
  163--172. Springer, 1997.

\bibitem{HosTakKanYaj97TCS}
K.~Hosaka, Y.~Takenaga, T.~Kaneda, and S.~Yajima.
\newblock Size of ordered binary decision diagrams representing threshold
  functions.
\newblock {\em Theoretical Computer Science}, 180(1):47 -- 60, 1997.

\bibitem{IwaNouYaj98}
Kazuo Iwama, Mitsushi Nouzoe, and Shuzo Yajima.
\newblock Optimizing {OBDDs} is still intractable for monotone functions.
\newblock In {\em Proceedings of the 23rd International Symposium on
  Mathematical Foundations of Computer Science (MFCS'98)}, volume 1450 of {\em
  Lecture Notes in Computer Science}, pages 625--635. Springer, 1998.

\bibitem{KayLafMos07Book}
Phillip Kaye, Raymond Laflamme, and Michele Mosca.
\newblock {\em An Introduction to Quantum Computing}.
\newblock Oxford University Press, 2007.

\bibitem{KitSheVya02Book}
Alexei~Yu. Kitaev, Alexander~H. Shen, and Mikhail~N. Vyalyi.
\newblock {\em Classical and Quantum Computation}, volume~47 of {\em Graduate
  Studies in Mathematics}.
\newblock AMS, 2002.

\bibitem{Knu09Book}
Donald~E. Knuth.
\newblock {\em The Art of Computer Programming, Volume 4, Fascicle 1: Bitwise
  Tricks \& Techniques; Binary Decision Diagrams}.
\newblock Addison-Wesley Professional, 1 edition, March 2009.

\bibitem{Kot14STACS}
Robin Kothari.
\newblock An optimal quantum algorithm for the oracle identification problem.
\newblock In {\em Proceedings of the 31st International Symposium on
  Theoretical Aspects of Computer Science {(STACS} 2014)}, pages 482--493,
  2014.

\bibitem{LeGMag18PODC}
Fran\c{c}ois Le~Gall and Fr{\'e}d{\'e}ric Magniez.
\newblock Sublinear-time quantum computation of the diameter in congest
  networks.
\newblock In {\em Proceedings of the 2018 ACM Symposium on Principles of
  Distributed Computing}, PODC '18, pages 337--346, 2018.

\bibitem{Lee59BSTJ}
C.~Y. {Lee}.
\newblock Representation of switching circuits by binary-decision programs.
\newblock {\em The Bell System Technical Journal}, 38(4):985--999, July 1959.

\bibitem{MeiSlo94MFCS}
Christoph Meinel and Anna Slobodov{\'a}.
\newblock On the complexity of constructing optimal ordered binary decision
  diagrams.
\newblock In {\em Proceedings of 19th Mathematical Foundations of Computer
  Science (MFCS '94)}, volume 841 of {\em Lecture Notes in Computer Science},
  pages 515--524. Springer Berlin Heidelberg, 1994.

\bibitem{MeiThe98Book}
Christoph Meinel and Thorsten Theobald.
\newblock {\em Algorithms and Data Structures in VLSI Design: {OBDD} -
  Foundations and Applications}.
\newblock Springer, 1998.

\bibitem{Min93DAC}
{Shin-ichi} Minato.
\newblock Zero-suppressed {BDD}s for set manipulation in combinatorial
  problems.
\newblock In {\em Proceedins of the 30th ACM/IEEE Design Automation
  Conference}, pages 272--277, June 1993.

\bibitem{Min11SAT}
Shin{-}ichi Minato.
\newblock \emph{{\(\pi\)}}{DD}: {A} new decision diagram for efficient problem
  solving in permutation space.
\newblock In {\em Proceedings of the 14th International Conference on Theory
  and Applications of Satisfiability Testing ({SAT} 2011)}, volume 6695 of {\em
  Lecture Notes in Computer Science}, pages 90--104, 2011.

\bibitem{NieChu00Book}
Michael~A. Nielsen and Isaac~L. Chuang.
\newblock {\em Quantum Computation and Quantum Information}.
\newblock Cambridge University Press, 2000.

\bibitem{SawTakYaj94IEICE}
Hiroshi Sawada, Yasuhiko Takenaga, and Shuzo Yajima.
\newblock On the computational power of binary decision diagrams.
\newblock {\em IEICE Trans. Info. \& Syst., D}, 77(6):611--618, 1994.

\bibitem{SekImaTan95ISAAC}
Kyoko Sekine, Hiroshi Imai, and Seiichiro Tani.
\newblock Computing the {Tutte} polynomial of a graph of moderate size.
\newblock In {\em Proceedings of the Sixth International Symposium on
  Algorithms and Computation (ISAAC '95)}, volume 1004 of {\em Lecture Notes in
  Computer Science}, pages 224--233. Springer, 1995.

\bibitem{Sie02DAM}
Detlef Sieling.
\newblock The complexity of minimizing and learning {OBDDs} and {FBDDs}.
\newblock {\em Discrete Applied Mathematics}, 122(1):263 -- 282, 2002.

\bibitem{Sie02InfoComp}
Detlef Sieling.
\newblock The nonapproximability of {OBDD} minimization.
\newblock {\em Information and Computation}, 172(2):103 -- 138, 2002.

\bibitem{TakYaj00DAM}
Yasuhiko Takenaga and Shuzo Yajima.
\newblock Hardness of identifying the minimum ordered binary decision diagram.
\newblock {\em Discrete Applied Mathematics}, 107(1-3):191--201, 2000.

\bibitem{Tan20SWAT}
Seiichiro Tani.
\newblock Quantum algorithm for finding the optimal variable ordering for
  binary decision diagrams.
\newblock In {\em 17th Scandinavian Symposium and Workshops on Algorithm
  Theory, {SWAT} 2020}, volume 162 of {\em LIPIcs}, pages 36:1--36:19. Schloss
  Dagstuhl - Leibniz-Zentrum f{\"{u}}r Informatik, 2020.
\newblock Also available at arxiv:1909.12658.

\bibitem{TanHamYaj96IEICE}
Seiichiro Tani, Kiyoharu Hamaguchi, and Shuzo Yajima.
\newblock The complexity of the optimal variable ordering problems of a shared
  binary decision diagram.
\newblock {\em IEICE Transactions on Information and Systems},
  E79-D(4):271--281, 1996.
\newblock (Conference version is in {\it Proceedings of the Fourth
  International Symposium on Algorithms and Computation (ISAAC'93)}, vol. 2906,
  pp.389--398, Lecture Notes in Computer Science, Springer, 1993).

\bibitem{TanIma94ISAAC}
Seiichiro Tani and Hiroshi Imai.
\newblock A reordering operation for an ordered binary decision diagram and an
  extended framework for combinatorics of graphs.
\newblock In {\em Proceedings of the Fifth International Symposium on
  Algorithms and Computation (ISAAC'94)}, volume 834 of {\em Lecture Notes in
  Computer Science}, pages 575--583. {Springer-Verlag}, 1994.

\bibitem{Weg00Book}
Ingo Wegener.
\newblock {\em Branching Programs and Binary Decision Diagrams}.
\newblock SIAM Monographs on Discrete Mathematics and Applications. SIAM, 2000.

\end{thebibliography}

\pagebreak
\appendix
\section*{Appendix}

\section{Proofs of Lemmas}\label{appdx:ProofsOfLemmas}

\begin{proofof}{\autoref{lm:DPgeneral}}
We focus on the simplest case of $m=1$ and write just $I$ instead of $I_1$, since
it is straightforward to generalize the proof to the case of $m\ge 2$.
Let $\sizei$ and $\sizej$ be the cardinalities of $I$ and $J$, respectively.

To show the first equality, let $j^*\deq \pi_{\inner{I,J}}[\sizei+\sizej]$. 
By definition, we have
\begin{align*}
\mincost{\inner{I,J}}&=\sum_{i\in I\sqcup J}\cost{i}{f}{\pi_{\inner{I,J}}}
= \sum_{i\in I\sqcup J\setminus{j^*}}\cost{i}{f}{\pi_{\inner{I,J}}}+\cost{j^*}{f}{\pi_{\inner{I,J}}}.
\end{align*}
The first term is equal to $\mincost{\inner{I,J\setminus{j^*}}}$, since otherwise
there exists $\pi'\in \Pi(\inner{I,J})$ with 
$\pi'[\sizei+\sizej]=j^*$ and
$\pi'[\ell]\neq \pi_{\inner{I,J}}[\ell]$ for some $\ell\in [\sizei +\sizej-1]$
such that 
\begin{align*}
\sum_{i\in I\sqcup J}\cost{i}{f}{\pi'}&=
\sum_{i\in  I\sqcup J\setminus{j^*}}\cost{i}{f}{\pi'}+\cost{j^*}{f}{\pi'}\\
&< \sum_{i\in  I\sqcup J\setminus{j^*}}\cost{i}{f}{\pi_{\inner{I,J}}}+\cost{j^*}{f}{\pi_{\inner{I,J}}}
=\mincost{\inner{I,J}},
\end{align*}
which contradicts the definition of $\mincost{\inner{I,J}}$,
where we use $\cost{j^*}{f}{\pi'}=\cost{j^*}{f}{\pi_{{\inner{I,J}}}}$ by \autoref{lm:OBDDwidth}.
The remaining term $\cost{j^*}{f}{\pi_{{\inner{I,J}}}}$ is equal to
$\cost{j^*}{f}{\pi_{\inner{I, J\setminus j^*,j^*}}}$ by \autoref{lm:OBDDwidth}. 
Thus, the first equality in the statement of the lemma holds.
Since $\mincost{\inner{I,J\setminus j^*}}=\sum_{i\in I\sqcup J\setminus j^*}\cost{i}{f}{\pi_{\inner{I,J\setminus j^*}}}$ by the definition, and \autoref{lm:OBDDwidth} implies that $\cost{i}{f}{\pi_{\inner{I,J\setminus j^*}}}=\cost{i}{f}{\pi_{\inner{I,J\setminus j^*,j^*}}}$ for every $i\in I\sqcup J\setminus j^*$,
it holds that $\mincost{\inner{I,J\setminus j^*}}=\sum_{i\in I\sqcup J\setminus j^*}\cost{i}{f}{\pi_{\inner{I,J\setminus j^*,j^*}}}$.
Therefore, $\mincost{\inner{I,J\setminus j^*}}+\cost{j^*}{f}{\pi_{\inner{I,J\setminus j^*,j^*}}}=\sum_{i\in I\sqcup J}\cost{i}{f}{\pi_{\inner{I,J\setminus j^*,j^*}}}$. This implies that the second equality in the lemma.
\end{proofof}

\section{Two-Parameter Case}\label{appdx:TwoParameters}
To improve the complexity bound further,
we use \autoref{lm:divide-and-conquer} recursively.
Let $\sizek_1$ and $\sizek_2$ be  parameters fixed later 
such that $0<\sizek_1<\sizek_2<n$.
By applying the lemma once, we have
\[
\mincost{[n]}=
\min_{K_2\subset [n], \abs{K_2}=\sizek_2}
\Oset{
\mincost{K_2}
+\mincost{\inner{K_2,[n]\setminus K_2}}[[n]\setminus K_2]
}.
\]
Then, we apply the lemma again to 
$\mincost{K_2}$ to obtain
\[
\mincost{K_2}=
\min_{K_1\subset K_2\colon \abs{K_1}=\sizek_1}
\Oset{
\mincost{{K_1}}
+\mincost{\inner{K_1,K_2\setminus K_1}}[K_2\setminus K_1]
}.
\]

To find the optimal $K_1$  from among $\binom{\sizek_2}{\sizek_1}$ possibilities
and the optimal $K_2$ from among $\binom{n}{\sizek_2}$, we use the quantum minimum finding (\autoref{lm:QuantumMinimumFinding}).
As in the one-parameter case, we perform the classical preprocessing
that computes
$\FS{K_1}$ for all subsets $K_1\subseteq [n]$ of cardinality $\sizek_1$.

If we define $\sizek_1\deq \floor{\alpha_1 n}$ and $\sizek_2 \deq \floor{\alpha_2n}$,
assuming that $\alpha_1<1/3$, 
the total time complexity (up to a polynomial factor)
is
\begin{align}
T(n)&=\sum_{\ell=1}^{\floor{\alpha_1 n}} 2^{n-\ell}\cdot \binom{n}{\ell}+L_3(n)
\lessapprox  2^{(1-\alpha_1 )n+\bentropy{\alpha_1 }n}+L_3(n),\label{eq:sec.2.2:T}
\end{align}
where
\begin{align}
L_3(n)&=\sqrt{\binom{n}{\floor{\alpha_2 n}}}\Oset{L_2(n)+3^{(1-\alpha_2)n}}
\lessapprox 2^{\frac{1}{2}\bentropy{\alpha_2}n}(L_2(n)+2^{(1-\alpha_2)(\log_2 3)n}),\label{eq:sec.2.2:L3}\\
L_2(n)&=\sqrt{\binom{\floor{\alpha_2 n}}{\floor{\alpha_1 n}}}
\Oset{L_1(n)+2^{(1-\alpha_2)n}3^{(\alpha_2-\alpha_1)n}}
\lessapprox
2^{\frac{1}{2}\bentropy{\alpha_1/\alpha_2}\alpha_2n}(L_1(n)+2^{(1-\alpha_2)n+(\alpha_2-\alpha_1)(\log_2 3)n}),\label{eq:sec.2.2:L2}\\
L_1(n)&=O^\ast(1).\label{eq:sec.2.2:L1}
\end{align}
Intuitively, 
$L_3(n)$ represents the time 
required for producing
$\FS{\inner{K_2,[n]\setminus K_2}}$
such that
$\mincost{\inner{K_2,[n]\setminus K_2}}$
is minimum over all subsets $K_2 \ (\subset [n])$
of cardinality $\sizek_2$,
when the collection of $\FS{K_2}$ for all $K_2$
is available at the cost of $L_2(n)$;
$L_2(n)$ represents the time 
required for producing
$\FS{\inner{K_1,K_2\setminus K_1}}$
for fixed $K_2$
such that
$\mincost{\inner{K_1,K_2\setminus K_1}}$
is minimum over all subsets $K_1 \ (\subset K_2)$
of cardinality $\sizek_1$,
when the collection of $\FS{K_1}$ for all $K_1$ is available
at the cost of $L_1(n)$.
Note that,
by \autoref{lm:ClassicalComposition},
the time required for computing $\FS{\inner{K_1,K_2\setminus K_1,[n]\setminus K_2}}$ from $\FS{\inner{K_1,K_2\setminus K_1}}$ is $O^\ast(3^{(1-\alpha_2)n})$,
and the time required for computing
$\FS{\inner{K_1,K_2\setminus K_1}}$ from $\FS{K_1}$
is
$O^\ast(2^{(1-\alpha_2)n}3^{(\alpha_2-\alpha_1)n})$.

Since we are not interested in polynomial factors in $T(n)$,
we ignore them in the following analysis of this system.

From Eqs.~(\ref{eq:sec.2.2:L2}) and (\ref{eq:sec.2.2:L1}),
$L_2(n)$ is at most
$2^{\frac{1}{2}\bentropy{\alpha_1/\alpha_2}\alpha_2n}\cdot 2^{(1-\alpha_2)n+(\alpha_2-\alpha_1)(\log_2 3)n}$.
Suppose that $L_2(n)$ equals this upper bound, since our goal is to upper-bound the complexity.
To balance the two terms on the right-hand side of \autoref{eq:sec.2.2:L3}, we set
$ L_2(n)=2^{(1-\alpha_2)(\log_2 3)n}$,
which implies
\begin{equation}\label{eq:twice:(1)}
\frac{1}{2}\alpha_2  \bentropy{\alpha_1/\alpha_2}+
(1-\alpha_2) + (\alpha_2-\alpha_1)\log_23
=(1-\alpha_2)\log_2 3.
\end{equation}
On the condition that this holds, $L_3(n)$ is at most 
$2^{\frac{1}{2}\bentropy{\alpha_2}n}2^{(1-\alpha_2)(\log_2 3)n}$.
Suppose again that $L_3(n)$ equals this upper bound.
To balance the two terms on the right-hand side in \autoref{eq:sec.2.2:T}, we have $L_3(n)=2^{(1-\alpha_1 )n+\bentropy{\alpha_1 }n}$, implying that
\begin{equation}\label{eq:twice:(2)}
{(1-\alpha_1)+\bentropy{\alpha_1}}=\frac{1}{2}\bentropy{\alpha_2}+(1-\alpha_2)\log_2 3.
\end{equation}
By numerically solving Eqs.~(\ref{eq:twice:(1)}) and (\ref{eq:twice:(2)}), we have
$\alpha_1^\ast=0.192755$ and $\alpha_2^\ast=0.334571$.
Note that $\alpha_1^\ast$ is less than $1/3$, as we assumed.
For $\alpha_1=\alpha_1^\ast$ and $\alpha_2=\alpha_2^\ast$,
\[T(n)\lessapprox {2^{\Cset{(1-\alpha_1^\ast)+\bentropy{\alpha_1^\ast}}n }}=O^\ast(\gamma_2),\]
where $\gamma_2=2.85690< \gamma_1$.

\newpage
\section{Numerical Optimization Data}\label{appdx:NumericalData}
\begin{table}[tbhp]
\caption{Values of $\gamma_d$ and the corresponding $\alpha_i$'s of algorithm $\qalgorithm{d,\vec{\alpha}}$: Each value is written with six digits, but the actual calculation is done with 20-digit precision.}
\label{tab:data_divideandconquer}
\begin{center}
\begin{tabular}{|c||c||c|c|c|c|c|c|}\hline 
$d$ & $\gamma_d$ & $\alpha_1$ & $\alpha_2$ &  $\alpha_3$  &  $\alpha_4$ &  $\alpha_5$ &$\alpha_6$ \\\hline\hline 
1 & 2.97625 & 0.274862 &--- & --- & --- & --- &---\\\hline 
2 & 2.85690 & 0.192754 &0.334571 &--- & --- & --- &---\\\hline 
3 & 2.83925 & 0.184664 & 0.205128 & 0.342677 & --- & --- &---\\\hline 
4 & 2.83744 & 0.183859 & 0.186017 &0.206375&0.343503 & --- &---\\\hline 
5 & 2.83729 & 0.183795 & 0.183967 & 0.186125 &0.206474 &0.343569 & --- \\\hline 
6 & 2.83728 & 0.183791 & 0.183802 & 0.183974 & 0.186131 & 0.206480 &0.343573\\\hline
\end{tabular}
\end{center}
\end{table}

\begin{table}[tbhp]

\begin{center}
\caption{Values of $\beta_6^{(\ell)}$ and the corresponding $\gamma^{(\ell)}$ and $\alpha_1^{(\ell)},\alpha_2^{(\ell)},\alpha_3^{(\ell)},\alpha_4^{(\ell)},\alpha_5^{(\ell)},\alpha_6^{(\ell)}$ of algorithm~$\cqalgorithm{\Gamma^{(\ell)}}{6,\vec{\alpha}^{(\ell)}}$ for $\ell=0,1,\dots, 9$:  Each value is written with six digits, but the actual calculation is done with 20-digit precision. The value of $\gamma^{(\ell)}$ in the row for $\ell=0$ is set to 3, since the algorithm $\ccalg$ is used as the subroutine $\Gamma^{(0)}$ and its running time is $O^*(3^n)$. The value of $\gamma^{(\ell)}$ in each row for $\ell\ge 1$
is set to the value of $\beta_6^{(\ell-1)}$ in the row for $\ell-1$, since the
$\cqalgorithm{\Gamma^{(\ell -1)}}{6,\vec{\alpha}^{(\ell-1)}}$ is used as $\Gamma^{(\ell)}$.
}
\label{tab:data_composition}
\begin{tabular}{|c|c||c|c|c|c|c|c|c|}\hline 
$\ell$ & $\beta_6^{(\ell)}$ & $\gamma^{(\ell)}$ & $\alpha_1^{(\ell)}$ & $\alpha_2^{(\ell)}$ & $\alpha_3^{(\ell)}$ & $\alpha_4^{(\ell)}$ & $\alpha_5^{(\ell)}$ & $\alpha_6^{(\ell)}$ \\\hline \hline
0 & 2.83728 & 3.00000 & 0.183792 & 0.183802 & 0.183974 & 0.186132 & 0.206480 & 0.343573 \\\hline 
1  & 2.79364 & 2.83728 & 0.165753 & 0.165759 &0.165857 & 0.167339 & 0.183883 & 0.312741 \\\hline 
2 & 2.77981 & 2.79364 & 0.160487 & 0.160491 & 0.160574 & 0.16189 & 0.177376 & 0.303603 \\\hline 
3 & 2.77521 & 2.77981 & 0.158777 & 0.15878 & 0.158859 & 0.160124 & 0.175273 & 0.300622 \\\hline 
4 & 2.77366 & 2.77521 & 0.158203 & 0.158207 & 0.158284 & 0.159532 & 0.174568 & 0.299621 \\\hline
5 & 2.77313 & 2.77366 & 0.158009& 0.158013 & 0.158089 & 0.159332 & 0.174330 & 0.299282 \\\hline 
6 & 2.77295 & 2.77313 & 0.157943 & 0.157947 & 0.158023 & 0.159264 & 0.174249 & 0.299166 \\\hline 
7 & 2.77289 & 2.77295 & 0.157920 & 0.157924 & 0.158000 & 0.159241 & 0.174221 & 0.299127 \\\hline 
8 & 2.77287 & 2.77289 & 0.157913 & 0.157916 & 0.157992 & 0.159233 & 0.174212 & 0.299114 \\\hline 
9 & 2.77286 & 2.77287 & 0.157910 &0.157914 & 0.157990 & 0.159230 & 0.174208 & 0.299109 \\\hline 
\end{tabular}
\end{center}
\end{table}

\end{document}